\documentclass[journal]{IEEEtran}
\ifCLASSINFOpdf
\else
   \usepackage{graphicx}
   \graphicspath{{../eps/}}
   \DeclareGraphicsExtensions{.eps}
\fi
\usepackage{fancyhdr}
\usepackage{amsmath,amssymb,amstext}
\usepackage{subeqnarray}
\usepackage{cite}
\usepackage{hhline}
\usepackage{bm}
\usepackage{soul}
\usepackage{slashbox}
\usepackage{times,amsmath,amsfonts}
\usepackage{subfigure}
\usepackage{amsmath,amssymb,bm}
\usepackage{graphicx,picture}
\usepackage{cite}
\usepackage{subfigure}
\usepackage{subeqnarray}
\usepackage{cases}
\usepackage{color}
\usepackage{overpic}
\usepackage{multirow}
\usepackage{booktabs}
\usepackage{epstopdf}
\usepackage{algorithmic}
\usepackage[ruled, linesnumbered]{algorithm2e}

\definecolor{r}{rgb}{1,0,0}
\definecolor{b}{rgb}{0,0,1}
\definecolor{k}{rgb}{0,1,1}

\usepackage{upgreek}

\newcounter{saveeqn}%

\allowdisplaybreaks
\DeclareMathSymbol{\Phi}{\mathord}{letters}{8}

\begin{document}
\title{Integrated Sensing and Channel Estimation by Exploiting Dual Timescales for Delay-Doppler Alignment Modulation}

\author{
\IEEEauthorblockN{Zhiqiang Xiao,~\IEEEmembership{Graduated Student Member, IEEE}, Yong Zeng,~\IEEEmembership{Senior Member, IEEE}, Fuxi Wen,~\IEEEmembership{Senior Member, IEEE}, Zaichen Zhang,~\IEEEmembership{Senior Member, IEEE}, and Derrick Wing Kwan Ng,~\IEEEmembership{Fellow, IEEE} }

\thanks{
Part of this work has been presented at the IEEE ICC 2023, Rome, Italy, 28 May - 01 Jun. 2023 \cite{xiao2023Exploiting}.

This work was supported by the National Key R\&D Program of China with grant number 2019YFB1803400,  by the Natural Science Foundation of China under grant 62071114, by the Fundamental Research Funds for the Central Universities of China under grant 3204002004A2 and 2242022k30005.

Z. Xiao, Y. Zeng, and Z. Zhang are with the National Mobile Communications Research Laboratory and Frontiers Science Center for Mobile Information Communication and Security, Southeast University, Nanjing 210096, China, and are also with the Purple Mountain Laboratories, Nanjing 211111, China (e-mail: \{zhiqiang\_xiao, yong\_zeng, zczhang\}@seu.edu.cn). (\emph{Corresponding author: Yong Zeng.})

F. Wen is with the School of Vehicle and Mobility, Tsinghua University, Beijing 100084, China (e-mail: wenfuxi@tsinghua.edu.cn).

D. W. K. Ng is with the University of New South Wales, Sydney, NSW 2052, Australia (e-mail: w.k.ng@unsw.edu.au).
}

}
\maketitle

\begin{abstract}
For integrated sensing and communication (ISAC) systems, the channel information essential for  communication and sensing tasks fluctuates across different timescales.
Specifically, wireless sensing primarily focuses on acquiring path state information (PSI) (e.g., delay, angle, and Doppler) of individual multi-path components to sense the environment, which usually evolves much more slowly than the composite channel state information (CSI) required for communications.
Typically, the CSI is approximately unchanged during the {\it channel coherence time}, which characterizes the statistical properties of wireless communication channels.
However, this concept is less appropriate for describing that for wireless sensing.
To this end, in this paper, we introduce a new timescale to study the variation of the PSI from a channel geometric perspective, termed {\it path invariant time}, during which the PSI largely remains constant.
Our analysis indicates that the path invariant time considerably exceeds the channel coherence time.
Thus, capitalizing on these dual timescales of the wireless channel, in this paper, we propose a novel ISAC framework exploiting the recently proposed delay-Doppler alignment modulation (DDAM) technique.
Different from most existing studies on DDAM or delay alignment modulation (DAM) that assume the availability of perfect PSI, in this work, we propose a novel algorithm, termed as adaptive simultaneously orthogonal matching pursuit with support refinement (ASOMP-SR), for joint environment sensing and PSI estimation.
We also analyze the performance of DDAM with imperfectly sensed PSI.
Simulation results unveil that the proposed ASOMP-SR algorithm outperforms the conventional orthogonal matching pursuit (OMP) in sensing.
In addition, DDAM-based ISAC can achieve superior spectral efficiency and a reduced peak-to-average power ratio (PAPR) compared to standard orthogonal frequency division multiplexing (OFDM).

\end{abstract}


\section{Introduction}
Integrated sensing and communication (ISAC) has been recognized as one of the six major usage scenarios for IMT-2030 \cite{ITU-R}.
It is poised to serve as a foundational infrastructure for enabling future networks, providing various emerging new services beyond communications.
This approach bridges the gap between physical and cyber worlds, paving the way for real-time digital twins to become a reality \cite{tong20216g,you2021towards,xiao2022overview}.
To realize ISAC systems, mono-static and bi-static stand out as two principle architectures.
Specifically, for cellular-based mono-static ISAC systems, either the communication base station (BS) or user equipment (UE) plays the dual role of a communication/sensing signal transmitter and the radar receiver \cite{xiao2022waveform}.
It requires that the BS or UE possess full-duplex capability, with communication and sensing tasks usually corresponding to two distinct wireless channels~\cite{liu2020joint,yuan2021integrated,pucci2022system}.
On the other hand, for bi-static ISAC systems, multi-path signals propagating between the involved communication BS and UE can be exploited for environment sensing.
In this case, communication and sensing may share a common wireless channel.
Furthermore, with the escalating scale of antennas and signal carrier frequency, wireless channels have become more sparse and exhibit more pronounced geometric characteristics than conventional sub-6 GHz channels \cite{rangan2014millimeter}.
As such, for bi-static ISAC systems, there is potential to establish a unified framework for both communication and sensing, pursuing a {\it mutualism} between them.

Bi-static ISAC systems have been studied in the literature for joint environment scatterer/target sensing and communication channel estimation.
For example, in \cite{liufan2020joint,huang2022joint,xu2023joint}, the authors considered that some of the radar targets also serve as the scatterers within the communication channel.
In this scenario, the BS first transmits downlink pilots for preliminary scatterers/targets detection.
Subsequently, the communication user sends some uplink pilots to assist the BS to pinpoint the scatterers/targets in the communication channel.
%Specifically, the authors in \cite{liufan2020joint} considered that the uplink pilots and radar echoes share partial angle-of-arrivals (AoAs).
%Hence, the communication channel estimation and radar targets sensing are executed based on uplink pilots and radar echoes, respectively.
%By contrast, the authors in \cite{huang2022joint} considered that the BS performs joint radar target detection and communication channel estimation, leveraging on both the radar echoes and received uplink pilots.
%Also, in \cite{xu2023joint}, the authors interested a broadband ISAC system and a location-centric channel model was proposed.
%Beyond the AoAs, it assumed that some of the radar targets and communication scatterers share common positions.
Besides, the authors in \cite{hong2022joint} considered a multi-hop propagation channel, where two geometry-based models were proposed to estimate the first and last hop scatterers across different bouncing orders.
In addition, the authors in \cite{fascista2021downlink} and \cite{wen20205g}
exploited the multi-path components for both position estimation and environment mapping.

Despite both communication and sensing experiencing a common wireless channel in the bi-static ISAC setup, their desired channel parameters usually alter across different timescales that have been overlooked in the literature, e.g., \cite{liufan2020joint,huang2022joint,xu2023joint,hong2022joint,fascista2021downlink,wen20205g}.
For communications, it is usually required to estimate the channel state information (CSI), which is approximately static during the {\it channel coherence time} that typically spans an order of millisecond (ms), depending on the channel mobility and carrier frequency \cite{tse2005fundamentals}.
By contrast, for wireless sensing,
it usually entails to estimate the angles, delays, and Doppler frequencies of each multi-path, which are highly related to the channel geometry and the moving states of scatterers/UEs.
In this paper, we refer to these as the {\it path state information} (PSI).
%Note that the PSI, especially the Doppler frequency, cannot be accurately estimated within a single channel coherence time, as this duration is too short to induce significant phase rotation.

Fortunately, practical PSI typically varies much more slowly than the CSI, say on the order of hundreds of ms compared with just tens of ms \cite{duel2007fading,rappaport2013millimeter,rangan2014millimeter}.
This is because the channel geometry associated with the scatterers/UEs alter relatively slowly compared to the composite CSI.
To analyze the variations of scatterers, a local scatter function (LSF) was derived in \cite{matz2005non}.
It describes the average power of scatterers in terms of delay and Doppler.
Additionally, a timescale, termed {\it stationarity time}, was defined from a statistical perspective for a non-wide-sense stationary uncorrelated scattering fading channel.
In particular, during this stationarity time, the LSF remains approximately constant~\cite{hlawatsch2011wireless}.
However, such a statistical definition may not be suitable to characterize the wireless sensing channel, which needs to be estimated deterministically.
Therefore, in addition to the channel coherence time, in this paper, we introduce a new timescale for the wireless channel from a geometry perspective, referred to {\it path invariant time}, during which the PSI remains essentially static.
Thus, the PSI can be estimated over this path invariant time.
There are several pieces of related literature that have studied the channel estimation across different timescales~\cite{guo2017millimeter,qin2018time,gao2015spatially}.
These methods leveraged the property that the angle information of multi-paths varies more slowly than the path gains for channel estimation.
However, these studies mainly focused on orthogonal frequency division multiplexing (OFDM) systems and only the angle invariant property was considered.
Thus, such channel estimation results cannot be directly applied to environment sensing.

In this paper, by exploiting the aforementioned dual timescales of the wireless channel for communication and sensing, we investigate a bi-static ISAC system exploiting the recently proposed delay-Doppler alignment modulation (DDAM) \cite{lu2023dd}.
Indeed, DDAM is an extension of the delay alignment modulation (DAM), which can transform a time-frequency double-selective fading channel into a simple additive white Gaussian noise (AWGN) channel \cite{lu2022delay,xiao2022integrated,lu2023delay2,lu2023ddris,ding2022channel}.
%This is achieved through delay-Doppler-angle processing, without requiring sophisticated channel equalization.
%Compared to OFDM, DDAM enjoys several advantages such as low peak-to-average power ratio (PAPR), high Doppler frequencies resilient, and  reduced guard interval (GI) overhead \cite{xiao2022integrated}.
Different from most existing studies on DDAM or DAM that assume perfect PSI \cite{lu2022delay,xiao2022integrated,lu2023delay2,lu2023ddris}, we consider a downlink bi-static ISAC system with DDAM that operates without any prior knowledge of the channel or wireless environment.
Compared to OFDM that usually requires the channel estimation in the time-frequency (TF) domain for transmit and receive beamforming, DDAM only requires the PSI.
As such, channel estimation for DDAM aligns with bi-static scatterer sensing.
This renders DDAM an appealing option for the seamless integration of environment sensing and channel estimation, fostering a symbiotic relationship between communication and sensing.
Furthermore, unlike in \cite{xiao2023Exploiting}, in this paper, we explicitly provide the definition of the path invariant time and a more challenging downlink bi-static ISAC case is considered.
The main contributions of this paper are summarized as follows.
\begin{itemize}
\item
First, we introduce a new timescale, termed path invariant time, for the wireless channel, during which the PSI is approximately unchanged.
Moreover, our analysis indicates that the path invariant time significantly exceeds the channel coherence time.
Building on this insight, we propose a novel framework for the bi-static ISAC system exploiting the DDAM technique, where the sensed PSI can be directly utilized for both DDAM-based signal transmission and environment sensing, thus ensuring a mutual benefit between communication and sensing.

\item
Second, we propose a novel channel estimation method for PSI sensing.
Specifically, as the PSI is approximately unchanged during the path invariant time,
we partition this period into two phases.
In Phase-I, the PSI is estimated and the estimation results are then employed in Phase-II for DDAM-based signal transmission, eliminating the need for additional pilots.
By exploiting the joint sparsity in the angular-delay domain, the angle-of-departures (AoDs) and delays of multi-paths are initially estimated.
To this end, we propose a novel algorithm, termed {\it adaptive simultaneous orthogonal matching pursuit with support refinement} (ASOMP-SR) that can more accurately estimate the number of multi-paths as well as the AoD and delay of each path with reduced pilot overheads, compared to the conventional orthogonal matching pursuit (OMP) algorithm.
After that, Doppler estimation is performed for each path.

\item
Third, the performance of DDAM-based ISAC with imperfectly sensed PSI is analyzed for both typical on-grid and off-grid scenarios.
Simulation results demonstrate that
DDAM-based ISAC can achieve higher communication spectral efficiency attributable to the reduced guard interval (GI) overhead and lower peak-to-average power ratio (PAPR) compared to OFDM.
\end{itemize}

The rest of this paper is organized as follows. Section~\ref{system model} introduces the considered bi-static ISAC system model, for which the dual timescales for communication and sensing are revealed.
The performance of DDAM-based communication with perfect PSI is derived in Section~\ref{DDAM based ISAC}.
After that, the signal processing of the proposed DDAM-based ISAC is discussed in Section~\ref{signal_DDAM}.
In Section~\ref{DDAM_commun}, considering imperfectly sensed PSI, the signal processing and performance analysis of DDAM-based communication are provided.
Simulation results and performance analysis are detailed in Section~\ref{simulation}.
Finally, we conclude the paper in Section~\ref{conclusion}.

{\it Notation}:
Scalars are denoted by the italic letters.
Vectors and matrices are denoted by boldface lower- and upper-case letters, respectively.
The inverse, pseudo-inverse, complex conjugate, transpose, and Hermitian transpose operations are given by $(\cdot)^{-1}$, $(\cdot)^{\dagger}$, $(\cdot)^{*}$, $(\cdot)^T$, and $(\cdot)^{H}$, respectively.
The $\iota_2$-norm is given by $\|\cdot\|$ and $\|\Gamma\|_c$ is the cardinality of the set $\Gamma$.
The support set of a vector $\mathbf{x}$ is given by $\mathrm{supp}\{\mathbf{x}\}$.
For a matrix $\mathbf{A}$, $[\mathbf{A}]_{{\Gamma},:}$ and $[\mathbf{A}]_{:,\Gamma}$ denote the sub-matrix of $\mathbf{A}$ whose column and row indices defined by $\Gamma$, respectively.
$\mathbb{C}^{M\times N}$ denotes the space of $M\times N$ complex-valued matrices.
The operator $\lceil\cdot\rceil$ is an integer ceiling operation, $\delta(\cdot)$ denotes the Kronecker delta function, $\star$ is the linear convolution, $\mathrm{Vec}(\cdot)$ represents the vectorization operation,
$i=\sqrt{-1}$ denotes the imaginary unit of complex numbers, and $\mathbb{E}(\cdot)$ denotes the statistical expectation.
The distribution of a circularly symmetric complex Gaussian (CSCG) random variable with zero mean and $\sigma^2$ variance is denoted by $\mathcal{CN}(0,\sigma^2)$, the uniform distribution over $[0,2\pi)$ is denoted as $\mathcal{U}(0,2\pi)$, and $\sim$ stands for ``distributed as''.
The modulo operation with respect to $X$ is represented by $\mathrm{mod}_X(\cdot)$, where $X$ is a positive integer.

\section{System model and Dual timescales}\label{system model}
\subsection{System Model}
\begin{figure} %system model Fig.1%
  \centering
  \includegraphics[width=0.48\textwidth]{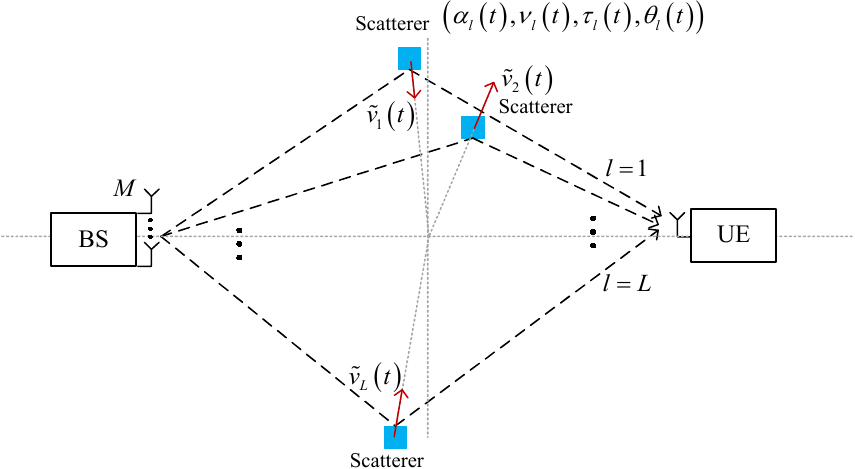}
  \caption{An illustration of the downlink bi-static ISAC system.}\label{system}\vspace{-0.3cm}
\end{figure}
As illustrated in Fig.~\ref{system}, we consider a cellular-based ISAC system, where a BS equipped with $M\gg 1$ transmit antennas and $B$ available bandwidth aims to serve a single-antenna UE\footnote{The current work can be easily extended to multiple UEs cases by transmitting orthogonal pilots for different UEs. Here, for ease of exposition, we consider the single UE case.} for downlink communication, while concurrently sense the wireless environment.
The baseband equivalent time-varying channel impulse response (CIR), $\mathbf{h}(t,\tau)\in\mathbb{C}^{M\times1}$, can be expressed as \cite{tse2005fundamentals}
\begin{equation}\label{channel model}
\begin{aligned}
\mathbf{h}^H(t,\tau)&=\sum\nolimits_{l=1}^{L(t)}\tilde{\alpha}_l(t)\psi(\tau-\tau_l(t))\mathbf{a}^H({\theta}_l(t)),
\end{aligned}
\end{equation}
where $t$ and $\tau$ denote the time and delay, respectively,
$\psi(\cdot)$ represents the band-limited pulse shaping response, $L(t)$ is the number of paths at time $t$, $\tau_l(t)$ denotes the delay of $l$th path, and
$\mathbf{a}({\theta}_l(t))\in\mathbb{C}^{M\times1}$ represents the transmit array response vector of the $l$th path with the AoD $\theta_l(t)$.
The complex-valued channel coefficient of the $l$th path can be further written as \cite{tse2005fundamentals}
\begin{equation}\label{path gain}
\tilde{\alpha}_l(t) = \frac{\beta_le^{-i2\pi f_c\tau_l(t)}}{d_l(t)},
\end{equation}
where $\beta_l>0$ is the amplitude factor at a reference distance, $f_c$ is the carrier frequency, $d_l(t)\triangleq d_l -\int_{-\infty}^{t}\tilde{v}_l(\zeta)\mathrm{d}\zeta$ represents the propagation distance of the $l$th path, with $d_l$ being the associated initial distance and $\tilde{v}_l(t)$ denoting the bi-static velocity as illustrated in Fig.~\ref{system}.
Note that the bi-static velocity $\tilde{v}_l(t)$ depends on the actual moving velocities and the relative locations of the scatterer and UE.
Specifically, a positive bi-static velocity means that the distance of the $l$th path decreases and vice versa \cite{kuschel2019tutorial}.
Obviously, the path delay $\tau_l(t)$ and the path distance $d_l(t)$ are related by
$\tau_l(t)=\frac{d_l(t)}{c}$, with $c$ denoting the speed of light.
Thus, by substituting \eqref{path gain} into \eqref{channel model}, we have
\begin{equation}\label{channel model2}
\begin{aligned}
\mathbf{h}^H(t,\tau) %&= \sum\limits_{l=1}^{L(t)}\frac{\beta_le^{-i2\pi f_c\tau_l(t)}}{d_l(t)}\psi(\tau-\tau_l(t))\mathbf{a}^H(\theta_l(t))\\
&=\sum\nolimits_{l=1}^{L(t)}\frac{\beta_le^{-i2\pi f_c\frac{d_l(t)}{c}}}{d_l(t)}\psi(\tau-\tau_l(t))\mathbf{a}^H(\theta_l(t))\\
%&=\sum\limits_{l=1}^{L(t)}\frac{\beta_le^{-i2\pi\left(d_l -\int_{-\infty}^{t}\tilde{v}_l(\zeta)\mathrm{d}\zeta\right)/\lambda}}{d_l(t)}\psi(\tau-\tau_l(t))\mathbf{a}^H(\theta_l(t))\\
%&=\sum\limits_{l=1}^{L(t)}\frac{\beta_le^{-i2\pi d_l/\lambda}}{d_l(t)}e^{i2\pi \int_{-\infty}^{t}\frac{\tilde{v}_l(\zeta)}{\lambda}\mathrm{d}\zeta}\psi(\tau-\tau_l(t))\mathbf{a}^H(\theta_l(t))\\
&=\sum\nolimits_{l=1}^{L(t)}\alpha_l(t)e^{i\varphi_l(t)}\psi(\tau-\tau_l(t))\mathbf{a}^H(\theta_l(t)),
\end{aligned}
\end{equation}
where $\alpha_l(t)\triangleq\frac{\beta_le^{-i2\pi d_l/\lambda}}{d_l(t)}$ is the complex-valued path gain of the $l$th path that excludes the Doppler effect, $\varphi_l(t)\triangleq 2\pi\int_{-\infty}^{t}\nu_l(\zeta)\mathrm{d}\zeta$ is the accumulated phase shift caused by the Doppler with the instantaneous Doppler frequency $\nu_l(t)\triangleq \frac{\tilde{v}_l(t)}{\lambda}$, and $\lambda=\frac{c}{f_c}$ is the signal wavelength.

For conventional TF domain modulation schemes such as OFDM, it requires to estimate the CIR $\mathbf{h}(t,\tau)$ or the CSI in the TF domain.
In contrast, for the recently proposed DDAM \cite{lu2023dd}, we only need to acquire the PSI, denoted by $\bm\eta(t)=\{\alpha_l(t),\nu_l(t),\tau_l(t), \theta_l(t)\}_{l=1}^{L(t)}$,
%including the number of multi-paths $L(t)$, the complex-valued path gain $\alpha_l(t)$, Doppler frequency $\nu_l(t)$, delay $\tau_l(t)$, and AoD $ \theta_l(t)$ of each path,
which is measured in the angular-delay-Doppler domain instead of the conventional TF domain.
Furthermore, as the PSI $\bm\eta(t)$ is intrinsically  linked to the physical propagation environment, the channel estimation for DDAM coincides perfectly with environment sensing.
On the one hand, the sensed PSI $\bm\eta(t)$ can be directly exploited to design the DDAM scheme; on the other hand, it can also be adopted to further determine the locations and motion state of the scatterers and UE.
This naturally paves the way for seamlessly integrating environment sensing and channel estimation for DDAM-based ISAC.

\subsection{Channel Coherence Time versus Path Invariant Time}
Despite the fact that both the CIR $\mathbf{h}(t,\tau)$ and the PSI $\bm\eta(t)$ are generally time-variant due to the motion of the scatterers and/or UE, they typically vary with different timescales.
Specifically, the CIR $\mathbf{h}(t,\tau)$ remains approximately unchanged during the channel coherence time $T_c$, which is defined from the statistical perspective and is approximately given by $T_c\approx \xi/\nu_{\max}$, where $\xi\le 1$ is a scaling factor \cite{cho2010mimo}, and $\nu_{\max}=\frac{\tilde{v}_{\max}}{\lambda}=\frac{\tilde{v}_{\max}f_c}{c}$ denotes the maximum Doppler frequency, with $\tilde{v}_{\max}$ being the maximum bi-static velocity.
Thus, as the carrier frequency $f_c$ and/or the motion velocity increases, the coherence time $T_c$ reduces such that the CIR $\mathbf{h}(t,\tau)$ varies more rapidly.
By contrast, the PSI $\bm\eta(t)$ alters much more slowly than the CIR $\mathbf{h}(t,\tau)$, since the geometry of the physical environment associated with the scatterers and/or UE varies relatively slowly.
Therefore, in addition to the conventional channel coherence time $T_c$, in this paper, we define a novel timescale, termed {\it path invariant time}.

{\it Definition 1}:
The path invariant time $\bar T$ is defined as the maximum time interval, during which the PSI $\bm\eta(t)$ remains approximately unchanged, i.e.,
\begin{equation}\label{PSI}
\bm\eta(t)\approx\bm\eta=\{{\alpha_l},\nu_l,\tau_l, \theta_l\}_{l=1}^L, t\in [t_0,t_0+\bar T).
\end{equation}

{\it Remark 1}:
Note that for a millimeter wave massive MIMO system with high temporal and spatial resolution, $\bar T$ is typically determined by the change rate of delays and AoDs.
Thus, to characterize $\bar T$ in \eqref{PSI}, we assume that over the time interval of interest, both the number of multi-paths $L(t)$ and the velocities of scatterers and UE are unchanged, denoted by $v_{s,1},\cdots,v_{s,L}$ and $v_u$, respectively.
Consequently, the instantaneous Doppler frequency $\nu_l(t)$ remains unchanged as well.
Furthermore, it is observed from \eqref{channel model2} that for $\alpha_l(t)$, only its amplitude varies over time via the path distance $d_l(t)$, which is much less sensitive than the variations in delay and AoD.
As a result, acquiring the path invariant time in \eqref{PSI} only requires  finding the maximum time interval $\bar T$ such that the changes of delays and AoDs of multi-paths over $[t_0,t_0+\bar T]$ can be considered negligible.
Therefore, we have the following theorem.

{\it Theorem 1}:
Denote by $\triangle \tau_{\max}(T)$ and $\triangle \bar\theta_{\max}( T)=\frac{1}{2}\left|\sin(\triangle\theta_{\max}(T))\right|$ the maximum possible variations of delay and normalized AoD over a time interval $T$, respectively,
with $\triangle\theta_{\max}(T)$ being the maximum AoD variation.
We have
\begin{equation}\label{delay angle max}
\begin{aligned}
&\triangle \tau_{\max}(T) \leq \frac{3v_{\max} T}{c},\\
&\triangle \bar\theta_{\max}(T) \leq \frac{v_{\max} T}{2(R_{\min}-v_{\max} T)},
\end{aligned}
\end{equation}
where $v_{\max}=\max\{v_{s,1},\cdots,v_{s,L},v_u\}$ denotes the maximum velocities of the scatterers and UE, and $R_{\min}=\min\{R_{s,1},\cdots,R_{s,L}\}$, with $R_{s,1},\cdots,R_{s,L}$ being the distances between the BS and the scatterers.
\begin{IEEEproof}
Please refer to  Appendix \ref{appendix a}.
\end{IEEEproof}

In practice, for an ISAC system with bandwidth $B$ and a uniform linear array (ULA) with $M$ transmit antennas and half-wavelength space, the system delay and normalized AoD resolutions are $1/B$ and $1/M$, respectively.
Thus, as long as
\begin{equation}\label{resolution}
\triangle\tau_{\max}(T)\le 1/B, \triangle\bar\theta_{\max}(T)\le 1/M,
\end{equation}
we can assume that the delays
and AoDs of multi-paths are approximately unchanged.
Thus, according to {\it Remark 1} and {\it Theorem 1}, we derive the following result:

{\it Theorem 2}:
For an ISAC system with bandwidth $B$ and $M\gg 1$ transmit antennas, the path invariant time $\bar T$ is
\begin{equation}\label{path invariant time0}
\begin{aligned}
\bar T &= \min\left\{\frac{c}{3v_{\max}B},\frac{2R_{\min}}{v_{\max}(M+2)}\right\}\\
&\overset{(a)}{\approx} \min\left\{\frac{c}{3v_{\max}B},\frac{2R_{\min}}{v_{\max}M}\right\},
\end{aligned}
\end{equation}
where the approximation $(a)$ holds when $M\gg 1$.
\begin{IEEEproof}
This result can be obtained by substituting \eqref{delay angle max} into \eqref{resolution}, which is omitted for brevity.
\end{IEEEproof}

In practice, for far-field communication systems, $R_{\min}\ge \frac{2D^2}{\lambda}$, with $D=\frac{\lambda}{2}M$ denoting the array aperture \cite{tse2005fundamentals}, the path invariant time $\bar T$ in \eqref{path invariant time0} satisfies
\begin{equation}\label{path invariant time2}
\bar T = \min\left\{\frac{c}{3v_{\max}B},\frac{2R_{\min}}{v_{\max}M}\right\}\ge \min\left\{\frac{c}{3v_{\max}B},\frac{\lambda M}{v_{\max}}\right\}.
\end{equation}

Note that from \eqref{path invariant time2}, we conclude that the path invariant time is typically much longer than the channel coherence time, i.e., $\bar T\gg T_c\approx\frac{\xi}{\nu_{\max}}=\frac{\xi\lambda}{\tilde{v}_{\max}}$, since both $\frac{c}{3v_{\max}B}=\frac{\lambda f_c}{3 v_{\max}B}\gg \frac{\xi\lambda}{\tilde{v}_{\max}}$ and $\frac{\lambda M}{v_{\max}}\gg \frac{\xi\lambda}{\tilde{v}_{\max}}$ hold with the typical relation $B\ll f_c$ and $M\gg 1$.
As an illustrative example, consider a mmWave ISAC system at a carrier frequency of $f_c=60$ GHz with $B=100$ MHz bandwidth and $M=128$ transmit antennas.
Assume that the locations of BS and UE are fixed while the scatterers move along the bi-static direction with the maximum velocity $\tilde{v}_{\max} = v_{\max}=50$ m/s and the minimum distance between the BS and scatterers is $R_{\min}=100$ m.
Thus, the corresponding maximum Doppler frequency is $\nu_{\max}=\frac{\tilde{v}_{\max}f_c}{c}=10$ kHz and the channel coherence time $T_c<\frac{1}{\nu_{\max}}=0.1$ ms.
On the other hand, according to \eqref{path invariant time0}, the path invariant time is $\bar T=20$ ms, which is much longer than the channel coherence time.

\subsection{Time-varying Channel with Dual Timescales}\label{double timescales}
\begin{figure} %system model Fig.1%
  \centering
  \includegraphics[width=0.38\textwidth]{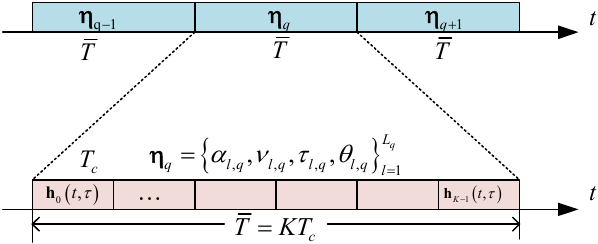}
  \caption{An illustration of the dual timescales for time-varying channels.
  %where $\bar T=KT_c$, $K\gg 1$. For each path invariant block $q$, $(q-1)\bar T\le t<q\bar T$, the PSI $\bm\eta(t)$ is approximately unchanged, i.e., $\bm\eta(t)\approx\bm\eta_q=\{\alpha_{l,q},\nu_{l,q},\tau_{l,q},\theta_{l,q}\}_{l=1}^{L_q}$, while for each channel coherence block $k$, $(k-1)T_c\le t < kT_c$, the CIR $\mathbf{h}_k(t,\tau)$ is approximately time-invariant, i.e., $\mathbf{h}_k(t,\tau)\approx\mathbf{h}_k(\tau)$.
  }\label{channel}\vspace{-0.3cm}
\end{figure}
By exploiting the above dual timescales, i.e., the path invariant time $\bar T$ and channel coherence time $T_c$, the time horizon $t$ can be partitioned into multiple blocks, as illustrated in Fig.~\ref{channel}.
For the $q$th path invariant block, i.e., $(q-1)\bar T\le t<q\bar T$, the PSI $\bm\eta(t)$ is approximately unchanged, i.e., $\bm\eta(t)\approx\bm\eta_q=\{\alpha_{l,q},\nu_{l,q},\tau_{l,q},\theta_{l,q}\}_{l=1}^{L_q}$, such that the CIR in \eqref{channel model2} reduces to
\begin{equation}\label{CIR_PIT}
\begin{aligned}
\mathbf{h}^H(t,\tau;q) = \sum\nolimits_{l=1}^{L_q}\alpha_{l,q}e^{i2\pi\nu_{l,q}t}\psi(\tau-\tau_{l,q})\mathbf{a}^H(\theta_{l,q}).
\end{aligned}
\end{equation}
Note that within each path invariant block, the CIR $\mathbf{h}(t,\tau;q)$ is a deterministic function of the PSI $\bm\eta_q$.
Thus, while the channel is still time-varying, as long as the PSI $\bm\eta_q$ is available, the CIR over the whole path invariant time $\bar T$ can be completely determined \cite{duel2007fading,wong2008sinusoidal}.
In the sequel, without loss of generality, we focus on one path invariant block and drop the index $q$ for ease of presentation.

Without loss of generality, each path invariant block can be further divided into $K$ consecutive channel coherence blocks, with $\bar T\triangleq KT_c$, where $K\gg1$ as $\bar T\gg T_c$.
For $0\le t< KT_c$, \eqref{CIR_PIT} can be further expressed as
\begin{equation}\label{CIR_PIT2}
\begin{aligned}
\mathbf{h}^H(t,\tau)
%&= \sum\limits_{l=1}^{L}\alpha_le^{i2\pi\nu_lt}\psi(\tau-\tau_l)\mathbf{a}^H(\bar\theta_l)\\
%&=\sum\limits_{k=0}^{K-1}\sum\limits_{l=1}^L\alpha_le^{i2\pi\nu_l(t-kT_c)}e^{i2\pi\nu_lkT_c}\psi(\tau-\tau_l)\mathbf{a}^H(\theta_l)\\
&=\sum\nolimits_{k=0}^{K-1}\mathbf{h}_k^H(t-kT_c,\tau),
\end{aligned}
\end{equation}
where
\begin{equation}\label{CIR_coherence0}
\mathbf{h}^H_k(t,\tau) =
\sum\limits_{l=1}^L\alpha_le^{i2\pi\nu_lt}e^{i2\pi\nu_lkT_c}\psi(\tau-\tau_l)\mathbf{a}^H(\theta_l), 0\le t< T_c,
\end{equation}
denotes the CIR over the $k$th channel coherence block.
Note that as $T_c\approx\xi/\nu_{\max}$ with $\xi\ll 1$, the variation of $\mathbf{h}_k^H(t,\tau)$ caused by the term $e^{i2\pi\nu_lt}$ in \eqref{CIR_coherence0} for $t\in[0,T_c)$ is negligible \cite{tse2005fundamentals}.
Thus, the CIR over each channel coherence block $k$ in \eqref{CIR_coherence0} is approximately invariant, i.e.,
\begin{equation}\label{CIR_coherence}
\small
\mathbf{h}^H_k(t,\tau) \approx \mathbf{h}^H_k(\tau)=\sum\nolimits_{l=1}^L\alpha_le^{i2\pi\nu_lkT_c}\psi(\tau-\tau_l)\mathbf{a}^H(\theta_l),
\end{equation}
which complies with the definition of the CIR in each channel coherence block $k$ time-invariant, while the variations of the CIR across different channel coherence block $k$ are captured by the term $e^{i2\pi\nu_lkT_c}$.

\section{DDAM-based ISAC}\label{DDAM based ISAC}
In this section, we first introduce the DDAM technique with perfect PSI and then explain why the DDAM technique can be utilized for ISAC by exploiting the aforementioned dual timescales of wireless channels.
\subsection{DDAM-based Communication with Perfect PSI}
For each path invariant block, $0\le t<\bar T = KT_c$, by sampling at $t=nT_s$ and $\tau=pT_s$, with $T_s=1/B$, the discrete-time equivalent of \eqref{CIR_PIT} is given by
\begin{equation}\label{CIR_dis}
\mathbf{h}^H[n,p] = \sum\nolimits_{l=1}^{L}\alpha_le^{i2\pi\nu_lnT_s}\psi(pT_s-\tau_l)\mathbf{a}^H(\theta_l),
\end{equation}
where $n=0,\cdots,KN-1$ with $T_c\triangleq NT_s$, $p=0,\cdots,P-1$, and $P\triangleq\lceil\tau_{\text{UB}}B\rceil$ denotes the maximum number of delay taps with $\tau_{\text{UB}}$ being a sufficiently large delay beyond which no significant power can be received.
To gain some insights, we first consider the {\it on-grid} case, i.e., the path delay $\tau_l$ is an integer multiple of the delay resolution $T_s$, i.e., $\tau_l=p_lT_s$, for some integer $p_l$.
Thus, $\psi(pT_s-\tau_l)$ in \eqref{CIR_dis} reduces to $\psi((p-p_l)T_s) = \delta(p-p_l)$.

Denote by $\mathbf{x}[n]$ the transmitted signal at time $n$.
With the discrete CIR in \eqref{CIR_dis}, the received signal at the UE is
\begin{equation}\label{rxSig}
\begin{aligned}
y[n] &= \mathbf{h}^H[n,p]\star\mathbf{x}[n] + z[n]\\
&= \sum\nolimits_{l=1}^L\alpha_le^{i2\pi\nu_lnT_s}\mathbf{a}^H(\theta_l)\mathbf{x}[n-p_l] + z[n],
\end{aligned}
\end{equation}
where $z[n]$ is the AWGN, satisfying $z[n]\sim\mathcal{CN}(0,\sigma^2)$, and $\sigma^2$ is the noise power at the receiver.

For DDAM, with perfect PSI $\bm\eta=\{\alpha_l,\nu_l,\tau_l,\theta_l\}_{l=1}^L$, the transmit signal is designed as \cite{lu2023dd}
\begin{equation}\label{DDAM_sig}
\mathbf{x}[n]=\sum\nolimits_{l=1}^L\mathbf{f}_ls[n-\kappa_l]e^{-i2\pi\nu_lnT_s},
\end{equation}
where $\kappa_l\triangleq p_{\max}-p_l$ is the delay pre-compensation, with $p_{\max}=\max\limits_{l=1,\cdots,L}p_l$, and $e^{-i2\pi\nu_lnT_s}$ is the Doppler pre-compensation, $\mathbf{f}_l\in\mathbb{C}^{M\times1}$ denotes the path-based beamforming vector, and $s[n]$ is the information bearing signal with normalized power $\mathbb{E}[|s[n]|^2]=1$.
%The transmit power of the DDAM signal is
%$\mathbb{E}[\|\mathbf{x}[n]\|^2]  = \sum\nolimits_{l=1}^L\|\mathbf{f}_l\|^2\le \bar P$,
%where $\bar P$ denotes the maximum average transmit power budget.

By substituting \eqref{DDAM_sig} into \eqref{rxSig}, we have
\begin{equation}\label{DDAMRx}
\small
\begin{aligned}
y[n]&=\sum\limits_{l=1}^L\sum\limits_{l'=1}^L{\alpha}_l{\mathbf{a}}^H(\theta_l)\mathbf{f}_{l'}s[n-\kappa_{l'}-p_l]e^{i2\pi(\nu_l-\nu_{l'})nT_s } + z[n]\\
&=\underbrace{\bigg(\sum\nolimits_{l=1}^{L}{\mathbf{h}}_l^H\mathbf{f}_l\bigg)s[n-p_{\max}]}_{\text{Time-invariant desired signal}}\\
&\quad+\underbrace{\sum\limits_{l=1}^{L}\sum\limits_{l'\neq l}^{L}{\mathbf{h}}_l^H\mathbf{f}_{l'}s[n-p_{\max}-\triangle p_{l,l'}]e^{i2\pi\triangle\nu_{l,l'}nT_s}}_{\text{Time-variant ISI}}+z[n],
\end{aligned}
\end{equation}
where ${\mathbf{h}}_l^H\triangleq{\alpha}_l{\mathbf{a}}^H(\theta_l)$.
The first term of \eqref{DDAMRx} is the desired signal, whose channel coefficient is time-invariant and fully utilizes the signal power of all multi-path components, while the second term is the ISI, which is introduced by the delay difference $\triangle p_{l,l'}\triangleq p_l-p_{l'}$, $\forall l\neq l'$ among different paths and is time-variant due to the Doppler difference $\triangle \nu_{l,l'}\triangleq \nu_l-\nu_{l'}$, $\forall l\neq l'$.

To mitigate the ISI, a path-based beamforming design was proposed \cite{lu2022delay}.
For instance, to completely eliminate the ISI, the path-based beamforming vectors $\{\mathbf{f}_l\}_{l=1}^L$ can be designed to ensure that ${\mathbf{h}}_l^H\mathbf{f}_{l'}=0$, $\forall l\neq l'$, which is referred to as the ISI-free zero-forcing (ZF) beamforming.
It is not difficult to verify that such a ZF condition is feasible as long as $M\ge L$.
In this case, \eqref{DDAMRx} reduces to
\begin{equation}\label{DDAMRx2}
y[n] = \left(\sum\nolimits_{l=1}^L{\mathbf{h}}_l^H\mathbf{f}_l\right)s[n-p_{\max}] + z[n].
\end{equation}
%where according to \cite{lu2022delay}, to mitigate the ISI, we can design $\mathbf{f}_l$, $l=1,\cdots,L$ as ${\mathbf{f}}_l=\frac{\sqrt{P_d}{\mathbf{Q}}_l{\mathbf{h}}_l}{\sqrt{\sum\nolimits_{l=1}^L\left\|{\mathbf{Q}}_l{\mathbf{h}}_l\right\|^2}}
%$, with ${\mathbf{Q}}_l=\mathbf{I}_M-{\mathbf{U}}_l({\mathbf{U}}_l^H{\mathbf{U}}_l)^{-1}{\mathbf{U}}_l^H$ and
%${\mathbf{U}}_l=[{\mathbf{h}}_1,\cdots,{\mathbf{h}}_{l-1},{\mathbf{h}}_{l+1},\cdots,{\mathbf{h}}_{{L}}]$.
It is observed from \eqref{DDAMRx2} that with DDAM, the received signal in \eqref{DDAMRx} that experiences the time-varying double-selective fading channel is efficiently transformed into \eqref{DDAMRx2} over an equivalent AWGN channel, without resorting to the complex channel equalization or multi-carrier transmission.

%\subsection{Why DDAM for ISAC}
\begin{figure} %system model Fig.1%
  \centering
  \subfigure[Conventional.]{
  \includegraphics[width=0.48\textwidth]{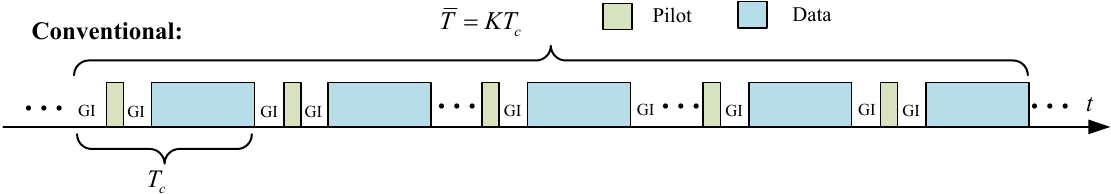}\label{block_con}
  }
  \\
  \subfigure[Proposed DDAM-based ISAC.]{
  \includegraphics[width=0.48\textwidth]{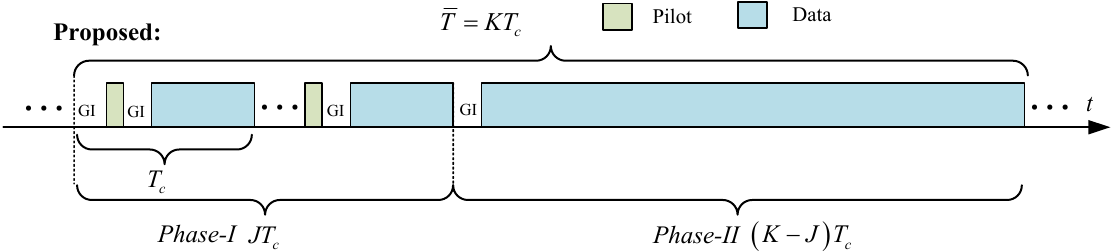}\label{block_pro}
  }
  \caption{Comparison between the conventional pilot-based communication channel estimation and the proposed DDAM-based ISAC.}\label{DDAM_ISAC}\vspace{-10pt}
\end{figure}

\subsection{DDAM-based ISAC with Dual Timescales}
For conventional schemes, such as OFDM, depicted in Fig.~\ref{block_con}, the channel estimation is performed over each channel coherence time $T_c$ to estimate the CSI in the TF domain.
Thus, pilots usually need to be inserted for each channel coherence block.
By contrast, for the proposed DDAM-based ISAC, the PSI ${\bm\eta}=\{{\alpha}_l,\nu_l,\tau_l,\theta_l\}_{l=1}^{L}$ is estimated in the angular-delay-Doppler domain.
Note that the PSI ${\bm\eta}$, especially the Doppler frequency $\nu_l$ cannot be accurately estimated over one single channel coherence time $T_c$, which  is too short to induce any notable phase rotation as evident from \eqref{CIR_coherence}.
By exploiting the path invariant property as discussed in Section~\ref{double timescales}, as shown
in Fig.~\ref{block_pro}, the entire path invariant block is divided into two phases, where the PSI ${\bm\eta}$ is estimated in Phase-I and then used for DDAM-based communication in Phase-II without requiring additional pilots and channel estimation.
Specifically, Phase-I consists of $J<K$ consecutive channel coherence blocks.
Within each block, angular-delay domain sensing is first performed at the UE and then the sensed angular-delay domain channel coefficients are reported to the BS via feedbacks.
Based on which, DAM-based communication is executed for each channel coherence block, without applying Doppler pre-compensation since it has not been sensed yet.
After that, the angular-delay domain coefficients sensed over $J$ channel coherence blocks are jointly exploited for Doppler estimation.
Finally, the complete PSI ${\bm\eta}=\{{\alpha}_l,\nu_l,\tau_l,\theta_l\}_{l=1}^{L}$ is sensed at the UE and then reported to the BS for DDAM-based communication in Phase-II.
Denote by $N_p$ and $N_g$ the lengths of pilot and GI, respectively.
As illustrated in Fig.~\ref{block_pro}, the total signalling overheads saving is $(K-J)(N_p+2N_g)-N_g$.
The mathematical details are discussed in the following.

\section{Concurrent Channel Estimation and Sensing for DDAM-Based ISAC}\label{signal_DDAM}
\subsection{Angular-delay Domain Sensing}\label{delay_angle_estimation}
According to \eqref{CIR_coherence}, for the $k$th channel coherence block, the time-invariant discrete-time CIR is
\begin{equation}\label{dis_CIR_k}
\mathbf{h}^H_k[n,p]\approx\mathbf{h}^H_k[p] = \sum\nolimits_{l=1}^L\alpha_{l,k}\psi(pT_s-\tau_l)\mathbf{a}^H(\theta_l),
\end{equation}
where $k=0,\cdots,K-1$, $p=0,\cdots,P-1$, and $\alpha_{l,k}\triangleq\alpha_le^{i2\pi\nu_lkT_c}$.
Denote by $\mathbf{p}_k[n]\in\mathbb{C}^{M\times1}$, $n=0,\cdots,N_p-1$, the transmitted pilots for the $k$th channel coherence block at time $n$, with the length of $N_p<N$ and channel training power $\mathbb{E}\left[\left\|\mathbf{p}_k[n]\right\|^2\right]=P_t$.
Let $\mathbf{p}_k[n]=\mathbf{0}_{M\times1}$ for $n<0$ or $n\ge N_p$ and set the length of GI as $N_g\ge P$ to avoid inter-block-interference (IBI) between the pilots and payload data, as shown in Fig.~\ref{block_pro}.
Note that at this stage, neither delay-Doppler compensation nor path-based beamforming is applied since the PSI is unavailable yet.

With the discrete-time CIR in \eqref{dis_CIR_k}, for the $k$th channel coherence block, the pilots received at the UE are
\begin{equation}\label{rxSig_est}
y_k[n] = \sum\limits_{p=0}^{P-1}\mathbf{h}_k^H[p]\mathbf{p}_k[n-p] + z[n], n=0,\cdots,\tilde{N}_p-1,
\end{equation}
where $\tilde{N}_p\triangleq N_p+P-1$.
It can be further rewritten as
\begin{equation}\label{rxSig_est2}
y_k[n] = \bar{\mathbf{h}}_k^H\bar{\mathbf{p}}_k[n] + z[n], n=0,\cdots,\tilde{N}_p-1,
\end{equation}
where $\bar{\mathbf{h}}_k\triangleq\mathrm{Vec}(\mathbf{H}_k)\in\mathbb{C}^{MP\times1}$, with $\mathbf{H}_k\triangleq[\mathbf{h}_k[0],\cdots,\mathbf{h}_k[P-1]]\in\mathbb{C}^{M\times P}$, and $\bar{\mathbf{p}}_k[n]\triangleq\mathrm{Vec}(\bar{\mathbf{P}}_k[n])\in\mathbb{C}^{MP\times1}$, with $\bar{\mathbf{P}}_k[n]\triangleq[\mathbf{p}_k[n],\cdots,\mathbf{p}_k[n-(P-1)]]\in\mathbb{C}^{M\times P}$.
By stacking the $\tilde{N}_p$ received signals in \eqref{rxSig_est2}, we have
\begin{equation}\label{rxSig_est3}
\mathbf{y}_k = \mathbf{P}_k^H\bar{\mathbf{h}}_k + \mathbf{z},
\end{equation}
where $\mathbf{y}_k=\big[y_k[0],\cdots,y_k[\tilde{N}_p-1]\big]^H\in\mathbb{C}^{\tilde{N}_p\times1}$,
$\mathbf{P}_k=\big[\bar{\mathbf{p}}_k[0],\cdots,\bar{\mathbf{p}}_k[\tilde{N}_p-1]\big]\in\mathbb{C}^{MP\times \tilde{N}_p}$, and $\mathbf{z}\in\mathbb{C}^{\tilde{N}_p\times1}$ is an AWGN vector.

To estimate the delays and spatial AoDs, we further express the channel matrix $\mathbf{H}_k$ in beamspace as \cite{bajwa2010compressed}
\begin{equation}\label{beamspace}
\mathbf{H}_k = \mathbf{A}\tilde{\mathbf{H}}_k,
\end{equation}
where $\mathbf{A}\in\mathbb{C}^{M\times M}$ denotes the beamspace transformation matrix, which depends on the geometrical structure of the transmit antenna array.
Here, we consider the ULA with half-wavelength antenna space.
In this case, $\mathbf{A}=\big[\tilde{\mathbf{a}}(0),\cdots,\tilde{\mathbf{a}}(M-1)\big]\in\mathbb{C}^{M\times M}$ becomes a discrete Fourier transform (DFT) matrix, with $\tilde{\mathbf{a}}(r)=\frac{1}{\sqrt{M}}\big[1,e^{i2\pi\frac{r-M/2}{M}},\cdots,e^{i2\pi\frac{(r-M/2)(M-1)}{M}}\big]^T$, $r=0,\cdots,M-1$.
Thus, $\tilde{\mathbf{H}}_k\triangleq\mathbf{A}^H\mathbf{H}_k\in\mathbb{C}^{M\times P}$ denotes the virtual channel matrix in the angular-delay domain, whose element at the $(r,p)$th angular-delay bin is
\begin{equation}\label{beamspace2}
\tilde{\mathbf{H}}_{k}[r,p]=\sum\limits_{l=1}^L\sqrt{M}\alpha_{l,k}^*\Gamma_M(\bar\theta_lM-(r-M/2))\psi(pT_s-\tau_l),\\
\end{equation}
where $\bar\theta_l\triangleq \frac{1}{2}\sin\theta_l$ with $\bar \theta_l\in[-\frac{1}{2},\frac{1}{2})$ for $\theta_l\in[-\frac{\pi}{2},\frac{\pi}{2})$, $r=0,\cdots,M-1$, $p=0,\cdots,P-1$, and $\Gamma_M(x)\triangleq\frac{\sin(\pi x)}{M\sin(\pi x/M)}e^{i\pi\frac{x(M-1)}{M}}$ is termed as the {\it Dirichlet} function \cite{bajwa2010compressed}.

Decompose that  $\tilde{\mathbf{H}}_{k}=\sum\nolimits_{l=1}^L\tilde{\mathbf{H}}_{l,k}$.
Then, the element of $\tilde{\mathbf{H}}_{l,k}$ at the $(r,p)$th angular-delay bin  is given by
\begin{equation}\label{beamspace_l}
\tilde{\mathbf{H}}_{l,k}[r,p]=\sqrt{M}\alpha_{l,k}^*\Gamma_M(\bar\theta_lM-(r-M/2))\psi(pT_s-\tau_l).\\
\end{equation}
In practice, the normalized AoDs $\bar\theta_l$ and delays $\tau_l$ are in general non-integer multiples of angular resolution $1/M$ and delay resolution $T_s=1/B$, which refers to as the {\it off-grid} case and may result in potential {\it power leakage} into the neighboring bins.
However, when the bandwidth $B$ and the number of transmit antennas $M$ are sufficiently large, we have the following properties:

\begin{figure} %system model Fig.1%
  \centering
  \includegraphics[width=0.38\textwidth]{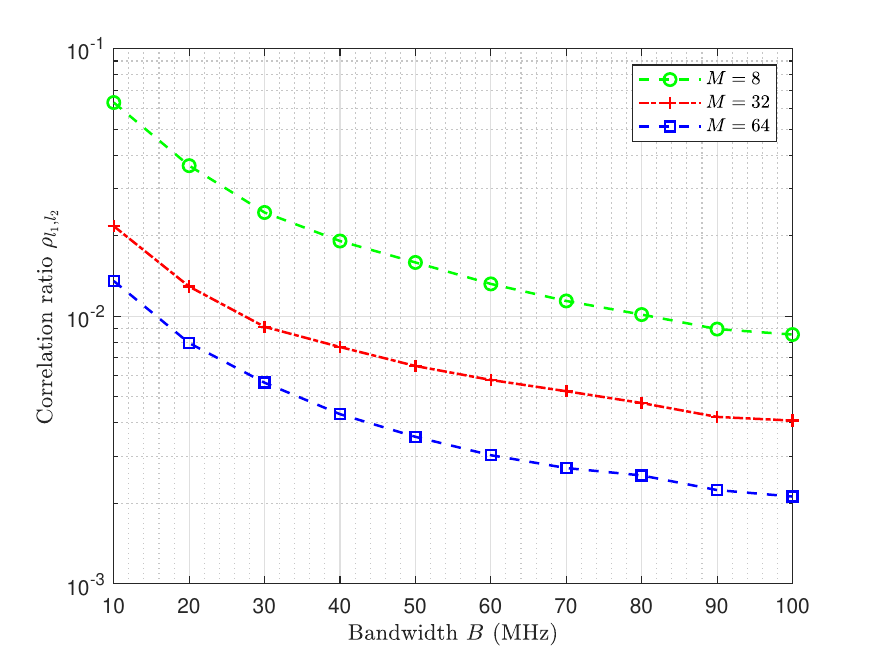}
  \caption{
  The correlation ratio versus bandwidth $B$ with different numbers of transmit antennas $M=8, 32$, and $64$.
  }\label{correlation}\vspace{-0.3cm}
\end{figure}
{\it Property 1}: Denote by $\tilde{\mathbf{h}}_{l,k}\triangleq\mathrm{Vec}(\tilde{\mathbf{H}}_{l,k})\in\mathbb{C}^{MP\times1}$ the path component of the $l$th path at the $k$th channel coherence block.
Define the cross-correlation factor between two different path components as
\begin{equation}\label{correlation_ratio}
\rho_{l_1,l_2}=\frac{\left|\tilde{\mathbf{h}}_{l_1,k}^H\tilde{\mathbf{h}}_{l_2,k}\right|}{\|\tilde{\mathbf{h}}_{l_1,k}\|\|\tilde{\mathbf{h}}_{l_2,k}\|}, \forall l_1\neq l_2.
\end{equation}
Note that as $\Gamma_M(\cdot)$ and $\psi(\cdot)$ in \eqref{beamspace_l} approach to the Kronecker delta function when the number of transmit antennas $M$ and bandwidth $B$ are sufficiently large, we have
$
\lim\limits_{M,B\rightarrow\infty}\rho_{l_1,l_2}=0, \forall l_1\neq l_2.
$
In Fig.~\ref{correlation}, the average correlation ratio $\rho_{l_1,l_2}$ versus the bandwidth $B$ with different numbers of transmit antennas $M$ over $10^4$ random channel realizations is shown.
It is observed that the correlation ratio $\rho_{l_1,l_2}$ decreases as the bandwidth $B$ and/or the number of transmit antennas $M$ increases.
Specifically, when $B=100$ MHz and $M=64$, $\rho_{l_1,l_2}$ is far below $10^{-2}$.
Thus, it implies that when the number of transmit  antennas $M$ and bandwidth $B$ are sufficiently large, $\tilde{\mathbf{H}}_{l,k}$,  $l=1,\cdots,L$ can be well separated in the angular-delay domain.

{\it Property 2}:
Since $\Gamma_M(\bar\theta_lM-(r-M/2))$ and $\psi(pT_s-\tau_l)$ have dominant power only if $r\approx \bar\theta_lM+M/2$ and $p\approx \tau_l/T_s$,  for each path $l$, we can approximate $\tilde{\mathbf{H}}_{l,k}$
by only retaining a few strongest bins of $\tilde{\mathbf{H}}_{l,k}$ and setting others to zero \cite{gao2017reliable}.
Thus, the support set of $\tilde{\mathbf{H}}_{l,k}$ is given by $\Omega_l\triangleq\{(r,p):r\in\mathcal{R}(r_l),p\in\mathcal{P}(p_l)\}$, with
\begin{equation}\label{region}
\begin{aligned}
&\mathcal{R}(r_l)\triangleq \mathrm{mod}_M\left\{r_l-\frac{V_r}{2},\cdots,r_l+\frac{V_r-2}{2}\right\},\\
&\mathcal{P}(p_l)\triangleq \mathrm{mod}_P\left\{p_l-\frac{V_p}{2},\cdots,p_l+\frac{V_p-2}{2}\right\},\\
\end{aligned}
\end{equation}
where $(r_l,p_l)$ corresponds to the angular-delay bin that has significant power of $\tilde{\mathbf{H}}_{l,k}$,
while $V_r$ and $V_p$ denote the number of neighboring bins of $(r_l,p_l)$ in angular and delay domains, respectively, the modulo operations $\mathrm{mod}_M(\cdot)$ and $\mathrm{mod}_P(\cdot)$ with respect to $M$ and $P$, respectively, guarantee that $\mathcal{R}(r_l)\subset\{0,\cdots,M-1\}$ and $\mathcal{P}(p_l)\subset\{0,\cdots,P-1\}$.

Therefore, the support set of  $\tilde{\mathbf{H}}_{k}=\sum\nolimits_{l=1}^L\tilde{\mathbf{H}}_{l,k}$ in the angular-delay domain is given by $\bar{\Omega}_k\triangleq\bigcup\nolimits_{l=1,\cdots,L}\Omega_l$, with the sparsity level of $|\bar\Omega_k|_c\triangleq \bar L_k$, where $L\le \bar L_k\ll MP$.
Thus, we can assume that $\tilde{\mathbf{H}}_k$ is sparse in the angular-delay domain.

By substituting \eqref{beamspace} into \eqref{rxSig_est3}, we have
\begin{equation}\label{rxSig_est4}
\begin{aligned}
\mathbf{y}_k&=\mathbf{P}_k^H\mathrm{Vec}(\mathbf{A}\tilde{\mathbf{H}}_k)+\mathbf{z}\\
&=\mathbf{P}_k^H(\mathbf{I}_{P}\otimes\mathbf{A})\tilde{\mathbf{h}}_k+\mathbf{z},
\end{aligned}
\end{equation}
where $\tilde{\mathbf{h}}_k\triangleq\mathrm{Vec}(\tilde{\mathbf{H}}_k)\in\mathbb{C}^{MP\times1}$ is a sparse vector that needs to be estimated and $\bar{\mathbf{h}}_k=(\mathbf{I}_P\otimes\mathbf{A})\tilde{\mathbf{h}}_k$ according to \eqref{rxSig_est3}.
Thus, to sense the delay and AoDs of multi-path components from $\mathbf{y}_k$ in \eqref{rxSig_est4}, we can formulate a sparse recovery problem as
\begin{equation}\label{rxSig_est5}
\mathbf{y}_k=\bm\Phi_k\tilde{\mathbf{h}}_k + \mathbf{z},
\end{equation}
where $\bm\Phi_k\triangleq\mathbf{P}_k^H(\mathbf{I}_P\otimes\mathbf{A})\in\mathbb{C}^{\tilde{N}_p\times MP}$ denotes the {\it sensing matrix} \cite{bajwa2010compressed}.
Note that as the PSI $\bm\eta$ remains static over the path invariant time $\bar T= KT_c$, $\tilde{\mathbf{H}}_k$ share a common sparsity structure $\forall k=0,\cdots,K-1$, i.e.,
\begin{equation}
{\Omega}_0={\Omega}_1=\cdots={\Omega}_{K-1}\triangleq\bar{\Omega},
\end{equation}
with the sparsity level $|\bar{\Omega}|_c\triangleq\bar L\ll MP$, but they have different coefficients, i.e., $\tilde{\mathbf{H}}_{k_1}[r,p]\neq\tilde{\mathbf{H}}_{k_2}[r,p]$, $\forall k_1\neq k_2$, $\forall (r,p)\in\bar{\Omega}$.
Thus, the recovery vectors $\tilde{\mathbf{h}}_k\triangleq\mathrm{Vec}(\tilde{\mathbf{H}}_k)$, $k=0,\cdots,K-1$, also have a common support set, i.e.,
\begin{equation}
\mathrm{supp}(\tilde{\mathbf{h}}_0)=\mathrm{supp}(\tilde{\mathbf{h}}_1)=\cdots=\mathrm{supp}(\tilde{\mathbf{h}}_{K-1})=\Theta,
\end{equation}
where $\Theta\triangleq\{g\mid g=pM+r,\forall (r,p)\in\bar \Omega\}$ with $|\Theta|_c\triangleq\bar L\ll MP$.
Therefore,
according to the distributed compressed sensing (DCS) theory \cite{sarvotham2005distributed}, the pilots over the first $J<K$ consecutive channel coherence blocks in Phase-I can be jointly employed to recover $\tilde{\mathbf{h}}_k$ from $\mathbf{y}_k$ in \eqref{rxSig_est5}, $k=0,\cdots,J-1$.

To this end, an $\ell_0$-norm minimization problem can be formulated as
\begin{equation}\label{op}
\small
\begin{aligned}
&\min\limits_{\{\tilde{\mathbf{h}}_k\}_{k=0}^{J-1}}&&\left(\sum\nolimits_{k=0}^{J-1}\left\|\tilde{\mathbf{h}}_k\right\|_{0}^2\right)^{1/2}\\
&\quad{\text{s.t.}} &&\sum\limits_{k=0}^{J-1}\left\|\mathbf{y}_k-\bm\Phi_k\tilde{\mathbf{h}}_k\right\|^2\le \epsilon,\\
& &&\mathrm{supp}(\tilde{\mathbf{h}}_0) = \mathrm{supp}(\tilde{\mathbf{h}}_1)=\cdots=\mathrm{supp}(\tilde{\mathbf{h}}_{J-1})=\Theta,
\end{aligned}
\end{equation}
where $\epsilon$ is a threshold related to the stop criterion.
To handle \eqref{op}, the simultaneous orthogonal matching pursuit (SOMP) algorithm is usually adopted \cite{sarvotham2005distributed}.
However, as a greedy pursuit algorithm, the recovery accuracy of SOMP cannot always be guaranteed if the stop criterion was not properly set \cite{do2008sparsity}.
To this end, we develop an adaptive SOMP with support refinement (ASOMP-SR) algorithm for angular-delay domain sensing, which is summarized in \textbf{Algorithm 1}.

\begin{algorithm}[t]
	\caption{Adaptive Simultaneous Orthogonal Matching Pursuit with Support Refinement (ASOMP-SR) }
	\label{alg1}
	\textbf{Initialization}:
    The number of coherence blocks $J^{(0)}=0$;
    Current coherence block index $u=1$;
    Residual power difference threshold $\epsilon_\mathrm{th}$;
	
	\Repeat{recovery difference $\vartheta_u$ does not decrease}
    {
        $J^{(u)} = J^{(u-1)} + 1$;

        Collect $\mathbf{y}_k$ and $\bm{\Phi}_k$ for $k=0,\cdots,J^{(u)}-1$;

        $\mathcal{I}^{(u)}=\emptyset$ \% Initialize the index set

        $\iota = 1$;  \% Iteration index

        $\mathbf{r}_k^{(0)}=\mathbf{y}_k\in\mathbb{C}^{\tilde{N}_p\times1}$, $k=0,\cdots,J^{(u)}-1$;

        \Repeat{residual difference $\epsilon_\iota\le \epsilon_\mathrm{th}$}
        {
            $\hat{g}_\iota=\arg\max\limits_{g=0,\cdots,MP-1}\sum\nolimits_{k=0}^{J^{(u)}-1}\left\|[\bm\Phi_k]_{:,g}^H\mathbf{r}_k^{(\iota-1)}\right\|$;

            $\mathcal{I}^{(u)} = \mathcal{I}^{(u)}\cup \hat{g}_\iota$;

            $\mathbf{r}_k^{(\iota)}=\mathbf{r}_k^{(\iota-1)}-[\bm\Phi_k]_{:,\mathcal{I}^{(u)}}[\bm\Phi_k]_{:,\mathcal{I}^{(u)}}^\dagger\mathbf{r}_k^{(\iota-1)}$, $k=0,\cdots,J^{(u)}-1$; \% Residual update

            $\epsilon_\iota = \frac{\sum\nolimits_{k=0}^{J^{(u)}-1}\left\|\mathbf{r}_k^{(\iota-1)}-\mathbf{r}_k^{(\iota)}\right\|^2}{\sum\nolimits_{k=0}^{J^{(u)}-1}\left\|\mathbf{r}_k^{(\iota)}\right\|^2}$
            \% Residual difference

            $\iota=\iota+1$;
        }
        $\bm\Psi_k^{(u)}=\left[\mathbf{0},\cdots,[\bm\Phi_k]_{:,\mathcal{I}^{(u)}}, \cdots,\mathbf{0}\right]\in\mathbb{C}^{\tilde{N}_p\times MP}, k=0,\cdots,J^{(u)}-1$;

        $\hat{\mathbf{h}}_k^{(u)}=(\bm\Psi_k^{(u)})^\dagger\mathbf{y}_k, k=0,\cdots,J^{(u)}-1$;

        Refine the common support set $\mathcal{I}^{(u)}$ and estimation results $\hat{\mathbf{h}}_k^{(u)}$, $k=0,\cdots,J^{(u)}-1$ (Algorithm 2).

        DAM-based communication for the current coherence block $u$. (Section~\ref{DDAM_commun}).

        \If{$u=1$}
        {
            $\vartheta_u=\infty$

        \Else{

            $\vartheta_u = \frac{\sum\nolimits_{k=0}^{J^{(u-1)}-1}\left\|\hat{\mathbf{h}}_k^{(u)}-\hat{\mathbf{h}}_k^{(u-1)}\right\|^2}{\sum\nolimits_{k=0}^{J^{(u-1)}-1}\left\|\hat{\mathbf{h}}_k^{(u)}\right\|^2}$;
            \% recovery  difference
            }
        }
        $u = u + 1$;
    }	
\end{algorithm}
Specifically, as shown from {\it Lines 3$\sim$ 16} of Algorithm 1, for the $u$th channel coherence block, the pilots received from the previous $J^{(u)}$ channel coherence blocks are jointly used for the common sparsity recovery based on the standard SOMP algorithm \cite{sarvotham2005distributed}, where the channel coefficient vectors over $J^{(u)}$ channel coherence blocks are jointly estimated as $\hat{\mathbf{h}}_k^{(u)}$, $k=0,\cdots,J^{(u)}-1$, as shown in {\it Line 16} of Algorithm~1.
Moreover, according to the DCS theory \cite{sarvotham2005distributed}, the probability of the exact recovery of the SOMP algorithm can be improved with the increasing of $J$.
Therefore, to guarantee the recovery accuracy of the proposed method, we introduce an outer loop to adaptively increase $J$ until the recovery difference  $\vartheta_u = \frac{\sum\nolimits_{k=0}^{J^{(u-1)}-1}\left\|\hat{\mathbf{h}}_k^{(u)}-\hat{\mathbf{h}}_k^{(u-1)}\right\|^2}{\sum\nolimits_{k=0}^{J^{(u-1)}-1}\left\|\hat{\mathbf{h}}_k^{(u)}\right\|^2}$ between two consecutive coherence blocks does not decrease, which guarantees the convergence of the proposed method.

\begin{algorithm}[t]
	\caption{Support Refinement (SR) }
	\label{alg1}
	\textbf{Input}: Estimated $\hat{\mathbf{h}}_k$, $k=0,\cdots,J-1$;
                     the number of neighboring bins $V_r$ and $V_p$;

    \textbf{Output}: $\check{\mathbf{h}}_{k}$, $k=0,\cdots,J-1$; $\hat{\Theta}_l$, $l=1,\cdots,\hat{L}$.
	
    \textbf{Initialization}: $\hat{L}=1$; $\hat{\Theta} = \emptyset$; $\mathbf{h}_t=\hat{\mathbf{h}}_0$
	
    \Repeat{power ratio $\xi$ doest not increase}
    {
         $\hat{g}_l = \arg\max\limits_{g=0,\cdots,MP-1} |{\mathbf{h}}_t[g]|$;

         $\hat{p}_l = \lfloor g_l /M \rfloor$;  \% Index of delay

         $\hat{r}_l = \hat{g}_l - \hat{p}_l M$;  \% Index of spatial AoD

         \% Detect the support set according to \eqref{region}

         $\mathcal{R}(\hat{r}_l)\triangleq \mod_M\{\hat{r}_l-\frac{V_r}{2},\cdots,\hat{r}_l+\frac{V_r-2}{2}\}$;

         $\mathcal{P}(\hat{p}_l)\triangleq \mod_P\{\hat{p}_l-\frac{V_p}{2},\cdots,\hat{p}_l+\frac{V_p-2}{2}\}$;

         $\hat\Omega_l = \{(r,p):r\in\mathcal{R}(\hat{r}_l), p\in\mathcal{P}(\hat{p}_l)\}$;

         $\hat{\Theta}_l\triangleq\{g:g= r + pM, (r,p)\in\hat{\Omega}_l \}$;

         $\hat{\Theta} = \hat{\Theta}\cup\hat{\Theta}_l$;

         \% Refine the estimation results

         $\check{\mathbf{h}}_{k} = \mathbf{0}_{MP\times1}; [\check{\mathbf{h}}_{k}]_{\hat{\Theta},:}= [\hat{\mathbf{h}}_k]_{\hat{\Theta},:}$, $k=0,\cdots,J-1$;

         $\xi = \frac{\sum\nolimits_{k=0}^{J-1}\|\check{\mathbf{h}}_{k}\|^2}{\sum\nolimits_{k=0}^{J-1}\|\hat{\mathbf{h}}_{k}\|^2}$; \% Calculate the power ratio

         $[\mathbf{h}_t]_{\hat{\Theta}_l,:} = 0$; \% Update residual channel vector

         $\hat{L} = \hat{L} + 1$;
    }	
\end{algorithm}

However, for the conventional SOMP algorithm \cite{sarvotham2005distributed}, the PSI ${\bm\eta}=\{{\alpha}_l,\nu_l,\tau_l,\theta_l\}_{l=1}^{L}$ cannot be accurately estimated when the sparsity level is unknown a prior and/or when the delays and normalized AoDs are non-integer multiples of resolutions.
For DDAM-based ISAC, to further sense the delay and angle information of multi-paths, after the channel coefficients have been estimated as $\hat{\mathbf{h}}_k$, $k=0,\cdots,J-1$\footnote{Here, for ease of presentation, the superscript $u$ is omitted.
}, support refinement (SR) is performed in {\it Line 17} of Algorithm 1 and the details are given in \textbf{Algorithm 2}.
Specifically, according to {\it Property 2} and \eqref{region}, the support set for each path $l$ in the angular-delay domain can be detected as $\hat\Omega_l = \{(r,p):r\in\mathcal{R}(\hat{r}_l), p\in\mathcal{P}(\hat{p}_l)\}$ as shown from {\it Lines 5$\sim$ 11} of Algorithm 2, as long as the strongest angular-delay bin of the $l$th path is estimated as $(\hat{r}_l,\hat{p}_l)$.
After that, the estimated channel coefficient vectors $\hat{\mathbf{h}}_k$ can be refined as $\check{\mathbf{h}}_{k} \in\mathbb{C}^{MP\times1}$, $k=0,\cdots,J-1$, as shown in {\it Line 15} of {Algorithm} 2, and the support set is refined as $\hat{\Theta} = \bigcup\limits_{l=1,\cdots,\hat{L}}\hat{\Theta}_l$ with          $\hat{\Theta}_l\triangleq\{g:g= r + pM, (r,p)\in\hat{\Omega}_l \}$ and $\hat{L}$ denoting the estimated number of multi-paths.
{Algorithm} 2 terminates when the power ratio between the refined and original channel coefficient vectors, i.e., $\xi = \frac{\sum\nolimits_{k=0}^{J-1}\|\check{\mathbf{h}}_{k}\|^2}{\sum\nolimits_{k=0}^{J-1}\|\hat{\mathbf{h}}_{k}\|^2}$ does not increase, which guarantees the convergence of the proposed method.
Based on this, for each channel coherence block, the DAM-based communication is performed as shown in {\it Line 18} of Algorithm 1, which will be discussed in Section~\ref{DDAM_commun}.

\subsection{Doppler Sensing}\label{doppler_estimation}
After obtaining the angular-delay domain channel coefficient vectors $\check{\mathbf{h}}_k\in\mathbb{C}^{MP\times1}$, $k=0,\cdots,J-1$ in Phase-I, we have $\check{\mathbf{H}}=[\check{\mathbf{h}}_{0},\cdots,\check{\mathbf{h}}_{J-1}]\in\mathbb{C}^{MP\times J}$.
For each coherence block $k$, we can decompose $\check{\mathbf{h}}_k$ as $\check{\mathbf{h}}_{k}=\sum\nolimits_{l=1}^{\hat{L}}\check{\mathbf{h}}_{l,k}$, which can be obtained by firstly setting as $\check{\mathbf{h}}_{l,k}=\mathbf{0}_{MP\times1}$, and then let $[\check{\mathbf{h}}_{l,k}]_{\hat\Theta_l,:}=[\check{\mathbf{h}}_{k}]_{\hat\Theta_l,:}$.
With perfect sensing in the angular-delay domain, we can assume that $\check{\mathbf{h}}_{l,k}\approx\tilde{\mathbf{h}}_{l,k}\triangleq\mathrm{Vec}(\tilde{\mathbf{H}}_{l,k})\in\mathbb{C}^{MP\times1}$.
According to \eqref{beamspace_l}, for each path $l$, we have
\begin{equation}
\small
\begin{aligned}
&\tilde{\mathbf{H}}_{l,k}[r,p]=\sqrt{M}\alpha_{l,k}^*\Gamma_M(\bar\theta_lM-(r-M/2))\psi(pT_s-\tau_l)\\
&=\sqrt{M}\alpha_{l}^*e^{-i2\pi\nu_lkT_c}\Gamma_M(\bar\theta_lM-(r-M/2))\psi(pT_s-\tau_l),
\end{aligned}
\end{equation}
which means that the elements of $\check{\mathbf{h}}_{l,k}\approx \mathrm{Vec}(\tilde{\mathbf{H}}_{l,k})$ contains the same Doppler frequency $\nu_l$.
Therefore,
to sense the Doppler frequency of the $l$th path, we have
\begin{equation}
\begin{aligned}
\mathbf{u}_l^H&=\check{\mathbf{h}}_{l,0}^H\check{\mathbf{H}}\\
&=\left[\check{\mathbf{h}}_{l,0}^H\check{\mathbf{h}}_{0},\check{\mathbf{h}}_{l,0}^H\check{\mathbf{h}}_{1},\cdots,\check{\mathbf{h}}_{l,0}^H\check{\mathbf{h}}_{J-1}\right]\\
&=\left[\check{\mathbf{h}}_{l,0}^H\sum\limits_{l'=1}^{\hat{L}}\check{\mathbf{h}}_{l',0},\check{\mathbf{h}}_{l,0}^H\sum\limits_{l'=1}^{\hat{L}}\check{\mathbf{h}}_{l',1},\cdots,\check{\mathbf{h}}_{l,0}^H\sum\limits_{l'=1}^{\hat{L}}\check{\mathbf{h}}_{l',J-1}\right]\\
&\overset{(a)}{\approx}\left[\check{\mathbf{h}}_{l,0}^H\check{\mathbf{h}}_{l,0},\check{\mathbf{h}}_{l,0}^H\check{\mathbf{h}}_{l,1},\cdots,\check{\mathbf{h}}_{l,0}^H\check{\mathbf{h}}_{l,J-1}\right]\\
&=\varsigma_l\left[1,e^{-i2\pi\nu_lT_c},\cdots,e^{-i2\pi\nu_l(J-1)T_c}\right],
\end{aligned}
\end{equation}
where the approximation in $(a)$ holds according to {\it Property 1} and \eqref{correlation_ratio}, when the number of transmit antennas $M$ and bandwidth $B$ are sufficiently large, we have ${\left|\check{\mathbf{h}}_{l,k}^H\check{\mathbf{h}}_{l',k}\right|} \approx 0, \forall l\neq l'$, and
\begin{equation}
\small
\varsigma_l\triangleq \sum\limits_{r=\hat{r}_l-\frac{V_r}{2}}^{\hat{r}_l+\frac{V_r}{2}}\sum\limits_{p=\hat{p}_l-\frac{V_p}{2}}^{\hat{p}_l+\frac{V_p}{2}}M|\alpha_l|^2\left|\Gamma_M(\bar\theta_lM-(r-M/2))\psi(pT_s-\tau_l)\right|^2 \end{equation}
denotes the processed path gain of the $l$th path.

Thus, the Doppler frequency of the $l$th path can be estimated via the inverse DFT (IDFT) as
\begin{equation}
\hat{\omega}_l = \arg\max\limits_{\omega}\left|\mathbf{u}_l^H\mathbf{v}(\omega)\right|,
\end{equation}
where $\omega = -\frac{N_oJ}{2},\cdots,\frac{N_oJ}{2}-1$ with $N_o$ being a positive integer that denotes the sampling factor and $\mathbf{v}(\omega) \triangleq [1,e^{i2\pi\frac{\omega}{N_oJ}},\cdots,e^{i2\pi\frac{\omega(J-1)}{N_oJ}}]^T$.
Thus, the Doppler frequency of the $l$th path is estimated as $\hat{\nu}_l=\frac{\hat{\omega}_l}{N_oJT_c}$ with the resolution of $\frac{1}{JT_c}$.
Note that with a larger $N_o$, we have a finer search grid such that the Doppler frequency can be estimated more accurately.
With the sensed Doppler frequency $\hat{\nu}_l$, the channel coefficient vector for the $l$th path over the $k$th channel coherence block can be further decomposed as $\check{\mathbf{h}}_{l,k}=\hat{\bm{\alpha}}_le^{-i2\pi\hat{\nu}_lkT_c}$.

Therefore, the complete PSI can be sensed as
\begin{equation}\label{est_psi}
\hat{\bm\eta}=\{\hat{\alpha}_l,\hat{\nu}_l,\hat{\tau}_l,\hat{\theta}_l\}_{l=1}^{\hat{L}},
\end{equation}
where $\hat{\alpha}_l=\hat{\mathbf{H}}_l[\hat{r}_l,\hat{p}_l]$, with $\hat{\mathbf{H}}_l\triangleq\mathrm{Vec}^{-1}(\hat{\bm\alpha}_l)\in\mathbb{C}^{M\times P}$, $\hat{\tau}_l=\hat{p}_lT_s$, and $\hat{\theta}_l=\arcsin\left(\frac{2(\hat{r}_l-M/2)}{M}\right)$, according to \eqref{beamspace_l}.

\section{DDAM-based Communication with Sensed Path State Information}\label{DDAM_commun}
With the sensed PSI, the DDAM-based communication can be performed.
In Phase-I, for the $k$th channel coherence block, with the channel in \eqref{dis_CIR_k}, the received signal at the UE is
\begin{equation}
\small
\begin{aligned}
y_k[n] &= \sum\limits_{p=0}^{P-1}\mathbf{h}_k^H[p]\mathbf{x}[n-p] + z[n]\\
&=\sum\limits_{l=1}^{L}\sum\limits_{p=0}^{P-1}\alpha_{l,k}\psi(pT_s-\tau_l)\mathbf{a}^H(\theta_l)\mathbf{x}[n-p] + z[n],
\end{aligned}
\end{equation}
where the impact of the Doppler frequency is integrated into the complex-valued path gain  $\alpha_{l,k}=\alpha_le^{i2\pi\nu_lkT_c}$, which is approximately unchanged within each coherence time $T_c$.
By contrast, in Phase-II, with the channel in \eqref{CIR_dis} over the remaining $(K-J)$ channel coherence blocks, the received signal is
\begin{equation}\label{ddam_rx}
\small
\begin{aligned}
y[n] &= \sum\limits_{p=0}^{P-1}\mathbf{h}^H[n,p]\mathbf{x}[n-p]+z[n]\\
&
=\sum\limits_{l=1}^{L} \sum\limits_{p=0}^{P-1}\alpha_le^{i2\pi\nu_lnT_s}\psi(pT_s-\tau_l)\mathbf{a}^H(\theta_l)\mathbf{x}[n-p]+z[n],
\end{aligned}
\end{equation}
where the Doppler effects spans over several channel coherence blocks.
%To that end, with the key ideas of DDAM, for Phase-I, only the delays of multi-paths are compensated since the Doppler frequency of each path has not been estimated yet, which refers to as DAM-based communication \cite{lu2022delay}.
%By contrast, in Phase-II, with the complete sensed PSI \eqref{est_psi}, both the Doppler frequencies and delays of multi-paths can be compensated, and the DDAM-based communication is performed.
Here, we focus on the signal processing for DDAM-based communication in Phase-II, which includes the DAM-based counterpart in Phase-I as a special case.
In the following, we analyze the performance of DDAM-based communication with the estimated PSI for the  on-grid case, while the performance of the off-grid case is evaluated by numerical simulations in Section~\ref{simulation}.

For the on-grid case, we have $\tau_l=p_lT_s$ and $\bar\theta_l=\frac{r_l-M/2}{M}$, $l=1,\cdots,L$, for some integer $p_l$ and $r_l$ such that \eqref{ddam_rx} reduces to \eqref{rxSig}.
Therefore, according to \eqref{DDAM_sig}, with the estimated PSI $\hat{\bm\eta}=\{\hat{\alpha}_l,\hat{\nu}_l,\hat{p}_l,\hat{r}_l\}_{l=1}^{\hat{L}}$, the transmitted DDAM signal is designed as
\begin{equation}\label{DDAM_sig2}
\small
\mathbf{x}[n] = \sum\nolimits_{l=1}^{\hat{L}}\hat{\mathbf{f}}_ls[n-\hat{\kappa}_l]e^{-i2\pi\hat{\nu}_lnT_s},
\end{equation}
where $\hat{\kappa}_l\triangleq \hat{p}_{\max}-\hat{p}_l$ with $\hat{p}_{\max}=\max\limits_{l=1,\cdots,\hat{L}}\hat{p}_l$.
By substituting \eqref{DDAM_sig2} into \eqref{rxSig}, we have
\begin{equation}\label{ddam_rx2}
y[n]=\sum\limits_{l=1}^L\sum\limits_{l'=1}^{\hat{L}}\mathbf{h}_l^H\hat{\mathbf{f}}_{l'}s[n-\hat{\kappa}_{l'}-p_l]e^{i2\pi\triangle \nu_{l,l'}nT_s } + z[n],
\end{equation}
where $\hat{\mathbf{f}}_l$, $l=1,\cdots,\hat{L}$ can be designed according to different criterions such that minimum mean square error (MMSE), maximum-ratio transmission (MRT), or ZF \cite{lu2022delay}.
For example, with the path-based ZF beamforming, let $\hat{\mathbf{f}}_l=\frac{\sqrt{P_d}{\mathbf{Q}}_l\hat{\mathbf{h}}_l}{\sqrt{\sum\nolimits_{l=1}^L\left\|{\mathbf{Q}}_l{\mathbf{h}}_l\right\|^2}}
$, with $P_d$ being the data transmit power, $\hat{\mathbf{Q}}_l=\mathbf{I}_M-\hat{\mathbf{U}}_l(\hat{\mathbf{U}}_l^H\hat{\mathbf{U}}_l)^{-1}\hat{\mathbf{U}}_l^H$ and
$\hat{\mathbf{U}}_l=[\hat{\mathbf{h}}_1,\cdots,\hat{\mathbf{h}}_{l-1},\hat{\mathbf{h}}_{l+1},\cdots,\hat{\mathbf{h}}_{\hat{L}}]$, where $\hat{\mathbf{h}}_l=\hat{\alpha}_l\mathbf{a}^H(\hat{\mathbf{\theta}}_l)$.

To detect the information signal, the receiver treats the delay tap that has the strongest power as the desired signal component while others as interference.
Thus, the received signal in \eqref{ddam_rx2} is rewritten as
\begin{equation}\label{decode}
\small
\begin{aligned}
&y[n] = \left(\sum\limits_{(l,l')\in\mathcal{D}(\varrho_\star)}\mathbf{h}_l^H\hat{\mathbf{f}}_{l'}e^{i2\pi\triangle\nu_{l,l'} nT_s}\right)s[n-\varrho_\star]\\
&+\sum\limits_{\varrho=\varrho_{\min},\atop
\varrho\neq\varrho_\star}^{\varrho{\max}}\left(\sum\limits_{(l,l')\in\mathcal{D}(\varrho)}\mathbf{h}_l^H\hat{\mathbf{f}}_{l'}e^{i2\pi\triangle\nu_{l,l'} nT_s}\right)s[n-\varrho] + z[n],
\end{aligned}
\end{equation}
where $\mathcal{D}(\varrho) = \left\{(l,l')\mid\hat{\kappa}_{l'}+p_l=\varrho\right\}$, $\varrho = \varrho_{\min},\cdots,\varrho_{\max}$ with $\varrho_{\max} = \min\limits_{l'=1,\cdots,\hat{L},\atop l=1,\cdots,L}(\hat{\kappa}_{l'} + p_l)$ and $\varrho_{\max}=\max\limits_{l'=1,\cdots,\hat{L},\atop l=1,\cdots,L} (\hat{\kappa}_{l'}+p_l)$, and $\varrho_\star=\arg\max\limits_{\varrho}\left|\sum\nolimits_{(l,l')\in\mathcal{D}(\varrho)}\mathbf{h}_l^H\hat{\mathbf{f}}_{l'}\right|$.

Denote by $\triangle\nu_{\max}\triangleq\max\limits_{(l,l')\in\mathcal{D}(\varrho_\star)}|\triangle\nu_{l,l'} |$.
Therefore, as long as $\triangle\nu_{\max}(K-J)T_c\ll 1$, \eqref{decode} reduces to
\begin{equation}\label{decode_ddam}
\small
\begin{aligned}
&y[n] = \left(\sum\limits_{(l,l')\in\mathcal{D}(\varrho_\star)}\mathbf{h}_l^H\hat{\mathbf{f}}_{l'}\right)s[n-\varrho_\star]\\
&+\sum\limits_{\varrho=\varrho_{\min},\atop
\varrho\neq\varrho_\star}^{\varrho_{\max}}\left(\sum\limits_{(l,l')\in\mathcal{D}(\varrho)}\mathbf{h}_l^H\hat{\mathbf{f}}_{l'}e^{i2\pi\triangle\nu_{l,l'} nT_s}\right)s[n-\varrho] + z[n].
\end{aligned}
\end{equation}
Thus, for the worst case, i.e., all the interferences in the second term of \eqref{decode_ddam} are  added constructively, the minimum achievable SINR is given by
\begin{equation}\label{SINR}
\small
\bar\gamma = \frac{\left|\sum\limits_{(l,l')\in\mathcal{D}(\varrho)}\mathbf{h}_l^H\hat{\mathbf{f}}_{l'}\right|^2}{\sum\limits_{\varrho=\varrho_{\min},\varrho\neq\varrho_\star}^{\varrho_{\max}}\left|\sum\limits_{(l,l')\in\mathcal{D}(\varrho)}\mathbf{h}_l^H\hat{\mathbf{f}}_{l'}\right|^2+\sigma^2}.
\end{equation}
The minimum achievable spectral efficiency (ASE) for the proposed DDAM-based ISAC over the whole path invariant time $\bar T$ is
\begin{equation}\label{rate}
\small
\mathcal{R}_{\text{t}} = \frac{1}{KN}\left(N_d\sum\limits_{k=0}^{J-1}\log_2(1+\bar\gamma_k)+(K-J)N\log_2(1+\bar\gamma)\right),
\end{equation}
where $N_d\triangleq N-N_p-2N_g$ is the number of payload data symbols for each channel coherence block in Phase-I, as illustrated in Fig.~\ref{block_pro}, and $\bar\gamma_k$ is the resulting SINR of the $k$th channel coherence block in Phase-I, which can be derived similarly from \eqref{decode} to \eqref{SINR}.
Note that %the resulting SINR $\bar\gamma_k$ increases for each channel coherence block in Phase-I, as the channel can be estimated more accurately with increasing number of pilots for joint channel estimation as discussed in Section~\ref{delay_angle_estimation}.
%Moreover,
if $J\ll K$, from \eqref{rate}, we have $\mathcal{R}_t\approx\log_2(1+\bar\gamma)$, for which the loss of spectral efficiency due to the pilot and GI overheads is negligible.

\section{Simulation Results}\label{simulation}
In this section, simulation results are provided to evaluate the performance of the proposed DDAM-based ISAC.
The BS is equipped with $M=64$ antennas, the carrier frequency is $f_c = 30$ GHz, the total bandwidth is $B=100$ MHz, and the maximum Doppler frequency is $\nu_{\max} = 4$ kHz.
Thus, the corresponding channel coherence time is approximately $T_c\approx \sqrt{\frac{9}{16\pi\nu_{\max}^2}}\approx 0.1$ ms, which includes $N = T_cB = 10^4$ symbols.
The path invariant time is assumed to be $\bar T = 50$ ms.
Thus, each path invariant block includes $K=\bar T/T_c = 500$ channel coherence blocks.
The direct distance from the UE to the BS is  $R_{u}=100$ m.
The number of multi-paths is set as $L=5$, whose corresponding scatterers are distributed in the space with the AoDs $\theta_l\in[-60^{\circ},60^{\circ}]$ and the distances $R_{s,l}\in[10\ \text{m},100 \ \text{m}]$ to the BS.
Thus, the distance from the $l$th scatterer to the UE  is $R_{su,l}=\sqrt{R_{u}^2+R_{s,l}^2-2R_uR_{s,l}\cos(\theta_l)}$, the propagation distance of the $l$th path is $d_l=R_{s,l}+R_{su,l}$, and the propagation delay for the $l$th path is $\tau_l=d_l/c$.
The complex-valued path gain of the $l$th path is modeled as $\alpha_l=\frac{\beta_le^{-i2\pi d_l/\lambda}}{d_l}$, where $\beta_l=d_l\sqrt{\frac{\lambda^2\zeta_l}{(4\pi)^3R_{s,l}^2R_{su,l}^2}}$ according to the bi-static radar equation \cite{kuschel2019tutorial}.

According to the compressed sensing theory \cite{bajwa2010compressed}, to guarantee the sparse recovery performance, the sensing matrix $\bm{\Phi}$ should satisfy the restricted isometry property (RIP).
Therefore, the $m$th element of the pilot $\mathbf{p}_k[n]$ at time $n$ for the $k$th channel coherence block is given by
$[\mathbf{p}_k[n]]_{m} = \sqrt{\frac{P_t}{M}}e^{i\phi_{n,k,m}}$, where $\phi_{n,k,m}$ satisfies the independent identical uniform distribution of $\phi_{n,k,m}\sim\mathcal{U}[0,2\pi)$ \cite{gao2015spatially}.

%The transmit pilots are designed as $\mathbf{p}_k[n]=\sqrt{P_t}\mathbf{g}_k[n]s_k[n]$, $n=0,\cdots,N_p-1$, where $s_k[n]$ and $\mathbf{g}_k[n]\in\mathbb{C}^{M\times1}$ denote the pilot symbol and precoder vector for the $k$th channel coherence block at time $n$, respectively \cite{gao2015spatially}.
%According to the compressed sensing theory \cite{bajwa2010compressed}, to guarantee the sparse recovery performance, the sensing matrix $\bm{\Phi}$ should satisfy the restricted isometry property (RIP).
%Therefore, the pilots $s_k[n]$, $n=0,\cdots,N_p-1$ are set as equipropable binary phase shift keying (BPSK) symbols, while the element of procoder vector $\mathbf{g}_k[n]$ is set as $\frac{1}{\sqrt{M}}e^{i\phi}$ with the random variable satisfying the uniform distribution of $\phi\sim\mathcal{U}[0,2\pi)$.

\subsection{Sensing \& Channel Estimation Performance}
In this subsection, we first investigate the sensing and channel estimation performance of the proposed ASOMP-SR algorithm.
The conventional OMP method is presented as a benchmark \cite{tropp2007signal}.
The normalized mean square error (NMSE) of the channel estimation over the $J$ coherence blocks is selected as the evaluation criterion, i.e.,
\begin{equation}
\text{NMSE} = \frac{1}{J}\sum\limits_{k=0}^{J}\left[\frac{\|\hat{\mathbf{h}}_k-\tilde{\mathbf{h}}_k\|^2}{\|\tilde{\mathbf{h}}_k\|^2}\right].
\end{equation}

\begin{figure} %system model Fig.1%
  \centering
  \includegraphics[width=0.38\textwidth]{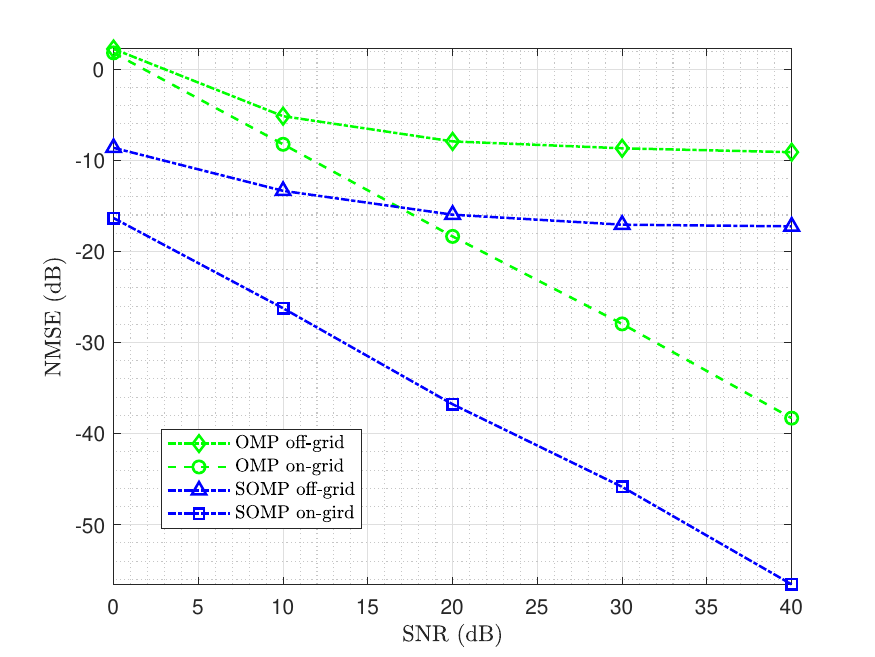}
  \caption{
  The NMSE performance of OMP and SOMP methods versus SNR for both the on-grid and off-grid cases.
  }\label{NMSE_on_off}\vspace{-0.3cm}
\end{figure}
Fig. \ref{NMSE_on_off} compares the NMSE of the OMP and SOMP methods against the SNR for both on-grid and off-grid cases.
According to \eqref{rxSig_est5}, the SNR is defined as $\mathrm{SNR}=\mathbb{E}\big[\|\bm\Phi_k\tilde{\mathbf{h}}_k\|^2\big]/\sigma^2$.
It can be observed that the NMSE decreases for all schemes with the increasing SNRs, as expected.
Moreover, the NMSE performance for the on-grid case is much better than that for off-grid case, due to the higher sparsity of the former.
Besides, for both  on-grid and off-grid cases, the SOMP method outperforms the OMP method, thanks to its exploitation of the joint sparsity of channel across multiple coherence blocks.

\begin{figure} %system model Fig.1%
  \centering
  \includegraphics[width=0.38\textwidth]{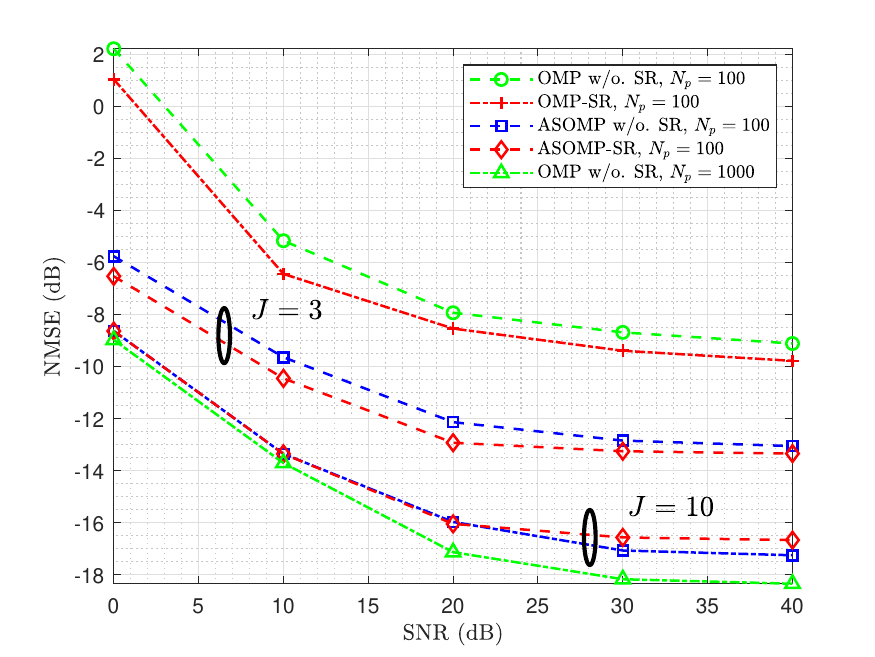}
  \caption{
  Comparison of the OMP and the proposed ASOMP-SR methods in terms of NMSE versus SNR for the off-grid case.
  }\label{NMSE_off}\vspace{-0.3cm}
\end{figure}
In Fig~\ref{NMSE_off}, we further compare the NMSE of OMP and the proposed ASOMP-SR methods in off-grid case.
It is observed that with the same amount of pilot overhead, i.e., $N_p = 100$, ASOMP can achieve better estimation performance with the increasing of $J$.
Specifically, when $J=10$, the ASOMP method with pilot overhead $N_p=100$ can achieve comparable estimation performance to the OMP method with $N_p=1000$.
This demonstrates that by leveraging the path invariant property and common sparsity across channel coherence blocks, the performance of channel estimation
and environment sensing can be significantly improved even with minimal pilot overhead.
For the proposed SR method, we retain $V_r=V_p=8$ strongest elements of each path component in the angular and delay domain, respectively.
With the proposed SR method, the NMSE performance of OMP and ASOMP method can be improved.
As for the ASOMP method, the performance gain diminishes with increasing $J$, since the corresponding NMSE performance tends to saturate as $J$ is sufficiently large.

%\begin{figure} %system model Fig.1%
%  \centering
%  % Requires \usepackage{graphicx}
%  \includegraphics[width=0.48\textwidth]{NMSEversusJ2.eps}
%  \caption{
%  The NMSE performance of ASOMP-SR method versus the number of coherence blocks $J$.
%  }\label{NMSE_J}\vspace{-0.3cm}
%\end{figure}
%To further explain the above observation, Fig.~\ref{NMSE_J} shows the NMSE of ASOMP-SR method as a function of $J$.
%It is observed that with the increasing of $J$, the NMSE of the proposed ASOMP method decreases and approaches to a limit when $J$ is large.
%Moreover, it is obtained that the proposed method SR can effectively improve the estimation performance when $J$  and/or SNR is small.
%Although such a performance gain of SR method is slight for NMSE, it will lead to significant gain for environment sensing.

\begin{figure} %system model Fig.1%
  \centering
  \subfigure[Environment sensing of OMP w/o. SR.]{
  \includegraphics[width=0.38\textwidth]{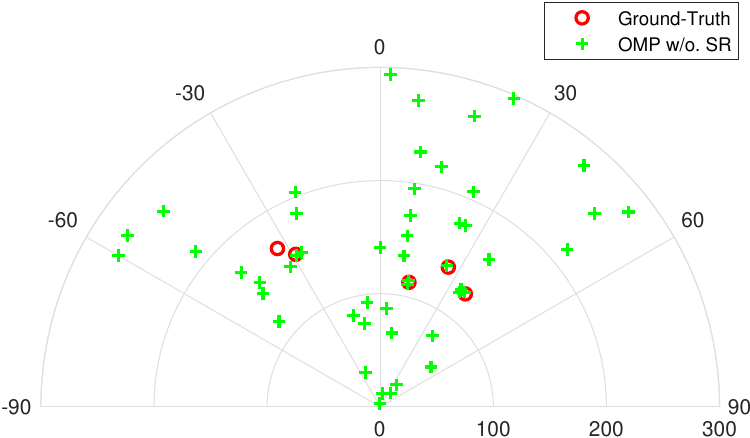}\label{omp_img}
  }
  \\
  \subfigure[Environment sensing of the proposed ASOMP-SR.]{
  \includegraphics[width=0.38\textwidth]{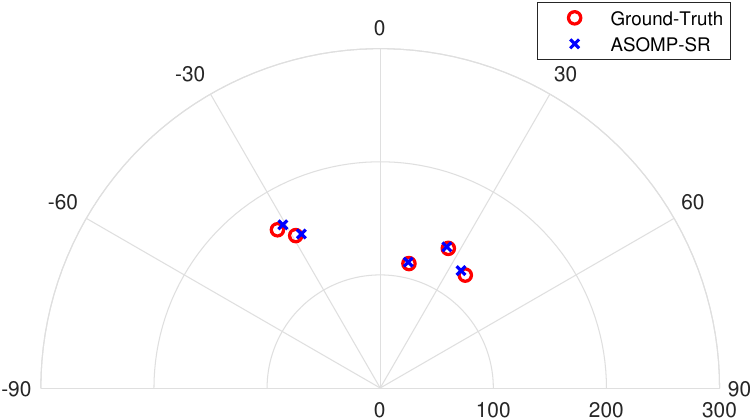}\label{somp_img}
  }
  \caption{Comparison of the conventional OMP without SR and the proposed ASOMP-SR in terms of   environment sensing performance in off-grid case.}\label{radarImg}\vspace{-10pt}
\end{figure}

Specifically, in Fig.~\ref{radarImg}, we compare the environment sensing performance of the conventional OMP method without SR to that of the proposed ASOMP-SR method in the off-grid scenario.
It is observed from Fig.~\ref{omp_img} that due to the power leakage phenomenon and the unknown sparsity level, the conventional OMP method without SR cannot accurately estimate PSI in terms of the number of paths $L$ and the delay and spatial AoDs of each multi-path.
By contrast, as shown in Fig.~\ref{somp_img}, the proposed ASOMP-SR method can accurately estimate such PSI for both sensing and DDAM-based communication.

\begin{figure} %system model Fig.1%
  \centering
  \includegraphics[width=0.38\textwidth]{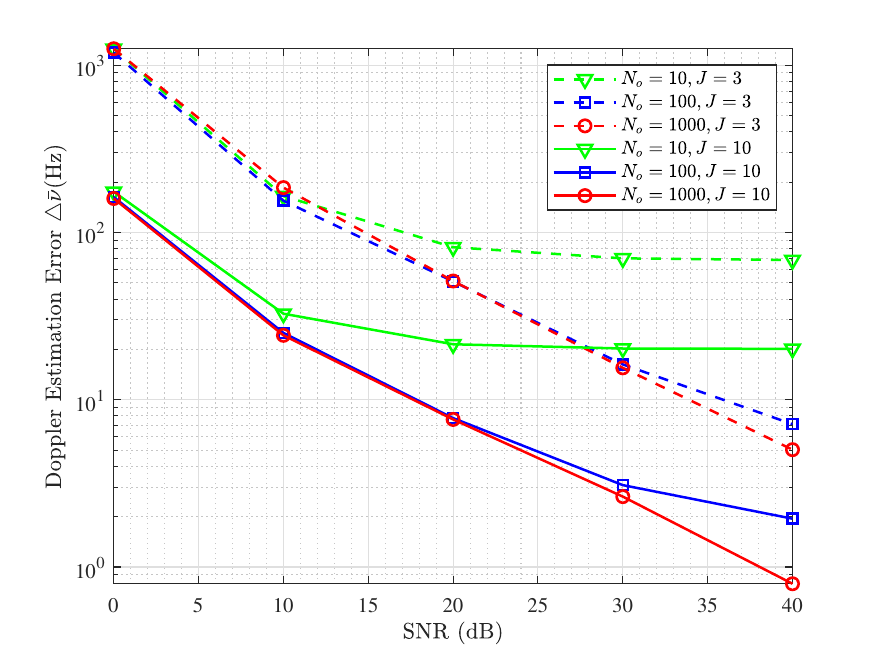}
  \caption{
  The Doppler estimation error versus SNR with different sampling factor $N_o=10, 100$, and $1000$.
  }\label{DopplerError}\vspace{-0.3cm}
\end{figure}
In Fig.~\ref{DopplerError}, we investigate the performance of the proposed Doppler frequency estimation method in Section~\ref{doppler_estimation}.
Here, we assume that the angular-delay domain channel is well estimated before performing Doppler estimation.
The average Doppler estimation error is defined as $\triangle\bar{\nu}\triangleq\left[\frac{1}{L}\sum\nolimits_{l=1}^{L}\left|\hat{\nu}_l-\nu_l\right|\right]$.
It is observed from Fig.~\ref{DopplerError} that $\triangle\bar{\nu}$ decreases with the increasing of SNR, the sampling factor $N_o$, and the number of coherence blocks $J$ in Phase-I, as expected.
Specifically, when $J=10$, $N_o=100$, and $\mathrm{SNR}=20$ dB, $\triangle\bar{\nu}$ drops below 10 Hz.
Thus, with the estimated Doppler frequency, for DDAM, as evident from \eqref{decode_ddam}, the time-varying channel in Phase-II can be effectively rendered time-invariant.

\subsection{Communication Performance}
\begin{figure} %system model Fig.1%
  \centering
  \subfigure[On-grid scenario.]{
  \includegraphics[width=0.38\textwidth]{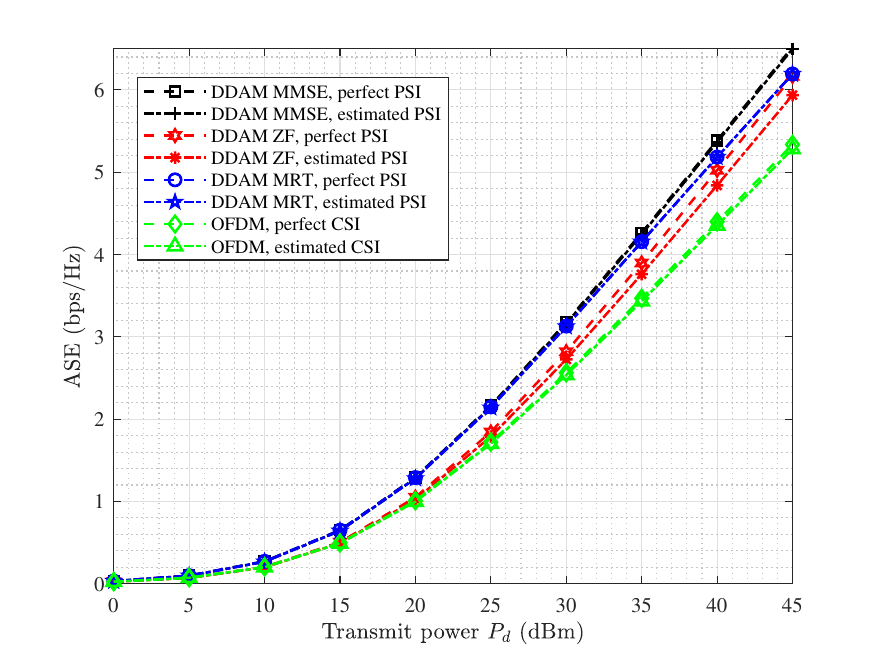}\label{Rate_on}
  }
  \\
  \subfigure[Off-grid scenario.]{
  \includegraphics[width=0.38\textwidth]{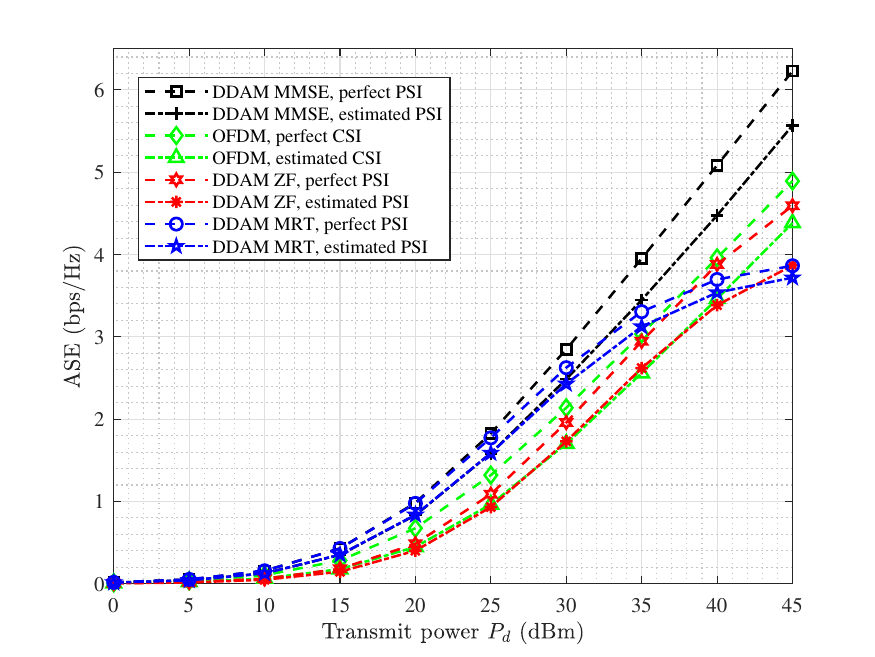}\label{Rate_off}
  }
  \caption{Comparison on the achievable communication rate versus transmit power $P_d$.}\label{RateVsPower}\vspace{-10pt}
\end{figure}

In this subsection, we evaluate the communication performance of the proposed DDAM-based ISAC system.
Fig.~\ref{RateVsPower} compares the achievable communication rate of the proposed DDAM-based ISAC system with the conventional OFDM for both on-grid and off-grid scenarios.
The noise power is set at $94$ dBm, while the transmit power increases from $0$ to $45$ dBm.
For an OFDM symbol with $W$ sub-carriers,
%for the $k$th channel coherence block with the CIR in \eqref{dis_CIR_k}, the channel of the ${w}$th sub-carrier is
%\begin{equation}\label{subcarrier}
%\mathbf{h}_{k,{w}}^H = \sum\limits_{p=0}^{P-1}\mathbf{h}_k^H[p]e^{-i\frac{2\pi {w}p}{W}}, {w}=0,\cdots,W-1,
%\end{equation}
the achievable communication rate is
\begin{equation}\label{rate_ofdm}
\small
\mathcal{R}_{\text{OFDM}} = \frac{N_d-N_{\text{OFDM}}N_{cp}}{N} \frac{1}{W}\sum\limits_{{w}=0}^{W-1}\log_2\left(1+\frac{|\mathbf{h}_{k,{w}}^H\mathbf{g}_{k,{w}}|^2}{\sigma^2/W}\right),
\end{equation}
where $\mathbf{h}_{k,w}\in\mathbb{C}^{M\times1}$ denotes the frequency domain channel of the $w$th  subcarrier over the $k$th channel coherence block,
$N_d\triangleq N-N_p-2N_g$ is total number of payload signal samples including the cyclic prefix (CP) for each channel coherence block, with $N_p$ and $N_g$ being the pilot and GI overhead, $N_{\text{OFDM}}=\frac{N_d}{W+N_{cp}}$ is the number of OFDM symbols for each channel coherence block, $N_{cp}$ is the length of CP, and $\mathbf{g}_{k,\mathrm{w}}\in\mathbb{C}^{M\times1}$ denotes the transmit beamforming vector for the ${w}$th subcarrier over the $k$th channel coherence block.
The length of GI or CP is set to the maximum delay tap, i.e., $N_g=N_{cp}=P=100$, and $W=512$.
For OFDM, the MRT beamforming is applied as $\mathbf{g}_{k,\mathrm{w}}=\sqrt{p_{\mathrm{w}}}\frac{\mathbf{h}_{k,\mathrm{w}}}{\|\mathbf{h}_{k,\mathrm{w}}\|}$, where $p_w$ is the transmit power of the $w$th sub-carrier channel and the water-filling power allocation is applied across the sub-carriers \cite{tse2005fundamentals}.
For comparison fairness, the channel estimation for OFDM is also performed based on the proposed ASOMP-SR method, where the PSI is initially estimated subsequently employed to reconstruct the sub-carrier channels.
The communication spectral efficiency of the proposed DDAM-based ISAC is given in \eqref{rate}, where the MMSE, ZF, and MRT path-based beamforming are applied~\cite{lu2022delay}.

Fig.~\ref{Rate_on} shows the results for the on-grid scenario.
It can be observed that the proposed DDAM-based ISAC can achieve higher communication spectral efficiency than that of OFDM.
This is expected since DDAM requires fewer pilot and GI overheads than OFDM, as evident form \eqref{rate} and \eqref{rate_ofdm}.
Moreover, by exploiting the path invariant property over $\bar T$, the proposed DDAM-based ISAC scheme requires less pilot overhead, where the PSI estimated in Phase-I can be directly employed for DDAM-based communication without incurring additional pilot and training overhead.

For the off-grid scenario shown in Fig.~\ref{Rate_off}, DDAM employing MRT or ZF beamforming performs worse than its on-grid counterpart.
This is because due to the potential power leakage across neighbor angular-delay bins, there may exist some channel coefficient vectors at different delay taps but corresponds to the same angular bin.
This makes it difficult to completely eliminate the ISI through path-based spatial beamforming.
However, the MMSE-based DDAM still outperforms OFDM significantly due to the reduced pilot and GI overheads.

\begin{figure} %system model Fig.1%
  \centering
  \includegraphics[width=0.38\textwidth]{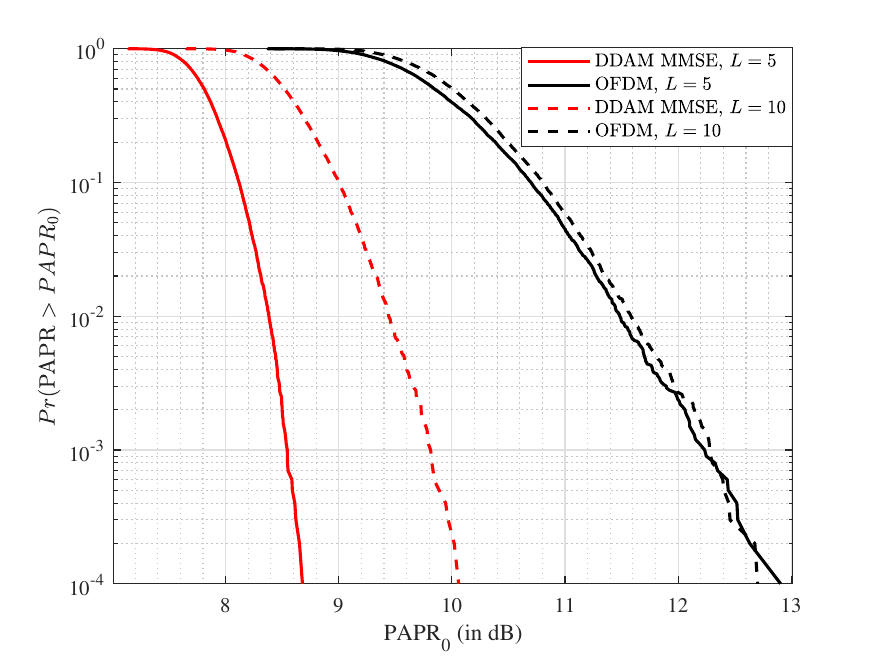}
  \caption{
  Comparison of the CCDFs of the PAPR between DDAM and conventional OFDM.
  }\label{PAPR}\vspace{-0.3cm}
\end{figure}
Fig.~\ref{PAPR} compares the complementary cumulative distribution function (CCDF) of PAPR for both DDAM and OFDM, with 16-QAM modulation.
It is observed from Fig.~\ref{PAPR} that DDAM can achieve much lower PAPR than OFDM for both scenarios with
$L=10$ to $L=20$ paths.
This is expected since the PAPR of DDAM or DAM is proportional to the number of paths, which is significant smaller than the number of sub-carriers for OFDM \cite{xiao2022integrated}.
Thus, DDAM-based ISAC can achieve superior power efficiency compared to its OFDM-based counterpart.

\section{Conclusion}\label{conclusion}
In this work, we proposed a novel timescale for wireless sensing, termed path invariant time, which was shown to be significantly longer than the channel coherence time for wireless communication.
Building upon this, we presented a novel DDAM-based ISAC scheme that capitalizes on the dual timescales of wireless channels.
This novel scheme unifies environment sensing and channel estimation for DDAM as a single task.
Based on the path invariant property and the common sparsity of the channel across channel coherence blocks, the DCS-based ASOMP-SR method was proposed to significantly reduce the required pilot overhead.
Simulation results demonstrated that the proposed DDAM-based ISAC can achieve superior communication spectral efficiency and lower PAPR than conventional OFDM rendering its appealing features for ISAC.

\begin{appendices}
\section{Proof of  Theorem 1}\label{appendix a}
\begin{figure} %system model Fig.1%
  \centering
  \includegraphics[width=0.38\textwidth]{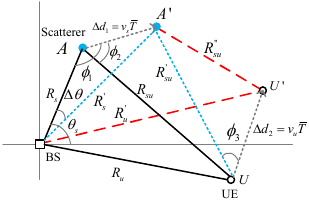}
  \caption{An illustration of the position variations of the scatterer and UE in a bi-static ISAC system.
  }\label{bistatic}\vspace{-0.3cm}
\end{figure}
As illustrated in Fig.~\ref{bistatic}, we consider the geometric among the BS, UE, and one scatterer, where the location of BS is fixed.
The initial locations of the scatterer and UE are denoted at ``$A$'' and ``$U$'', respectively, and their initial distances from the BS are denoted by $R_s$ and $R_u$, respectively.
Besides, the initial distance between the scatterer and UE is $R_{su}$.
The initial AoD from the BS to the scatterer is $\theta_s$.

First, we consider the case that the location of the UE is fixed, while the scatterer moves from ``$A$'' to ``$A'$'' during $[t_0,t_0+\bar T]$ with a velocity of $v_s$.
Thus, the distance from  ``$A$'' to ``$A'$'' is $\triangle d_1=v_{\mathrm{s}}\bar T$.
According to the law of cosines, the distance from the BS to  ``$A'$'' is
\begin{equation}\label{R1}
R_s' = \sqrt{R_s^2 + \triangle d_1^2 - 2R_s\triangle d_1\cos\phi_1}.
\end{equation}
Similarly, the distance from the scatterer at  ``$A'$'' to the UE at ``$U$'' is
\begin{equation}\label{R2}
R_{su}' = \sqrt{R_{su}^2 + \triangle d_1^2 - 2 R_{su}\triangle d_1\cos\phi_2}.
\end{equation}
Following that, by fixing the location of the scatterer at ``$A'$'' while the UE moving from ``$U$'' to ``$U'$'' with the velocity of $v_u$, the distance from ``$U$'' to ``$U'$'' is $\triangle d_2 = v_{u}\bar T$ and the distance between ``$A'$'' and ``$U'$'' is
\begin{equation}\label{R2.2}
R_{su}'' = \sqrt{R_{su}'^2 + \triangle d_2^2 - 2R_{su}'\triangle d_2\cos\phi_3}.
\end{equation}
Therefore, the total distance variation from the BS to the UE via the scatterer is
\begin{equation}\label{range variant}
|\triangle d| = |R_s' + R_{su}'' - (R_s+R_{su})|.
\end{equation}
In practice, we have $\triangle d_1,\triangle d_2\ll R_s, R_{su}$.
Thus, by applying the first-order Taylor series expansion to \eqref{R1}-\eqref{R2.2}, we have $R_s'\approx R_s - \triangle d_1\cos\phi_1$, $R_{su}'\approx R_{su} - \triangle d_1\cos\phi_2$, and $R_{su}''\approx R_{su}' - \triangle d_2\cos\phi_3\approx R_{su} - \triangle d_1\cos\phi_2 -\triangle d_2\cos\phi_3$.
Therefore, \eqref{range variant} is simplified as
\begin{equation}\label{v_r}
\begin{aligned}
|\triangle d| &\approx |\triangle d_1\cos\phi_1 +\triangle d_1\cos\phi_2+\triangle d_2\cos\phi_3|\\
&=|v_s\bar T(\cos\phi_1+\cos\phi_2)+v_u\bar T\cos\phi_3|\\
&\le (2v_s+v_u)\bar T.
\end{aligned}
\end{equation}

On the other hand, for the variation of the AoD $\triangle \theta$, it is only related to the position variation of the scatterer, while independent of the location of the UE.
As illustrated in Fig.~\ref{bistatic}, according to the law of cosines, we have
\begin{equation}\label{AoD}
\cos(\triangle \theta) = \frac{R_s^2 + R_s'^2 -\triangle d_1^2}{2R_sR_s'}.
\end{equation}
By substituting \eqref{R1} into \eqref{AoD}, it can be simplified as
\begin{equation}\label{v_theta}
\begin{aligned}
\cos(\triangle \theta) &=  \frac{R_s-\triangle d_1\cos\phi_1}{\sqrt{R_s^2 + \triangle d_1^2 - 2 R_s \triangle d_1\cos\phi_1}}.
\end{aligned}
\end{equation}
Thus, the normalized AoD variation is
\begin{equation}\label{angle_v}
\begin{aligned}
|\triangle\bar\theta|&=\frac{1}{2}\left|\sin(\triangle \theta)\right| = \frac{1}{2}\sqrt{1-\cos^2(\triangle \theta)}\\
&=\frac{\triangle d_1|\sin \phi_1|}{2\sqrt{R_s^2+\triangle d_1^2-2R_s \triangle d_1\cos\phi_1}}\le \frac{\triangle d_1}{2(R_s-\triangle d_1)}\\
&=\frac{v_\mathrm{s}\bar T}{2(R_s-v_\mathrm{s}\bar T)}.
\end{aligned}
\end{equation}
Denote by $v_{\max}=\max\{v_{s,1},\cdots,v_{s,L},v_u\}$ the maximum velocities of the scatterers/UE, with $v_{s,1},\cdots,v_{s,L}$ and $v_u$ being the velocities of the scatterers and UE, respectively, and $R_{\min}=\min\{R_{s,1},\cdots,R_{s,L}\}$ represents the minimum distance between the BS and the scatterers/UE.
According to \eqref{range variant} and \eqref{angle_v}, we have
\begin{equation}\label{max_variation}
\begin{aligned}
&\frac{|\triangle d|}{c} \le  \frac{3v_{\max}\bar T}{c}\triangleq\triangle \tau_{\max}(\bar T),\\
&|\triangle\bar\theta| \le \frac{v_{\max}\bar T}{2(R_{\min}-v_{\max}\bar T)}\triangleq\triangle\bar\theta_{\max}(\bar T).
\end{aligned}
\end{equation}
Thus, the proof of {\it Theorem 1} is completed.

\end{appendices}

\bibliographystyle{IEEEtran}
\bibliography{DoubleTimescales}

\end{document}